\documentclass[camera,letterpaper,nomarginnotes,nonarrowgutter,hyperref]{jpaper}
\usepackage{amsmath,amssymb,amsfonts}
\usepackage{algorithm}
\usepackage[noend]{algpseudocode}
\usepackage{algorithmicx}
\usepackage{graphicx}
\usepackage{textcomp}
\usepackage{xcolor}

\usepackage{makecell}
\usepackage{array}
\usepackage{booktabs}

\usepackage[sort,compress]{cite}

\usepackage{xcolor, soul}
\usepackage{xspace}
\usepackage{multirow}
\usepackage{tabularx}
\usepackage{tikz}
\usepackage[bottom]{footmisc}
\newcommand*\circled[1]{\tikz[baseline=(char.base)]{\node[shape=circle,fill,inner sep=0.5pt] (char) {\textcolor{white}{#1}};}}
\usepackage[most]{tcolorbox}
\usepackage{amsmath}

\usepackage{fancyhdr}
\usepackage{mathtools}

\usepackage{booktabs}

\usepackage{pgfplots}
\pgfplotsset{compat=1.17} %

\usepackage{stfloats}

\usepackage{enumitem}

\usepackage{mwe}

\usepackage[most]{tcolorbox}

\usepackage{pbox}

\definecolor{lavenderr}{rgb}{0.71, 0.49, 0.86}

\usepackage[most]{tcolorbox}   %
\tcbset{
  observebox/.style={
    colframe=lavenderr!90!black,      %
    colback=lavenderr!15,              %
    boxrule=1pt,                   %
    arc=1mm,                       %
    left=0.01mm, right=0.01mm,       %
    top=0.01mm, bottom=0.01mm,       %
    fonttitle=\bfseries,           %
    before skip=6pt, after skip=6pt
  }
}

\usepackage{wrapfig} 

\usepackage{setspace}

\usepackage[noend]{algpseudocode}
\usepackage{xcolor}
\usepackage{amsmath}

\definecolor{algokeyword}{RGB}{0, 0, 150} %
\definecolor{algocomment}{RGB}{0, 100, 0} %
\definecolor{algotext}{RGB}{50, 50, 50}   %

\algrenewcommand\algorithmicrequire{\textbf{\textcolor{algokeyword}{Input:}}}
\algrenewcommand\algorithmicensure{\textbf{\textcolor{algokeyword}{Output:}}}
\algrenewcommand\algorithmicif{\textbf{\textcolor{algokeyword}{if}}}
\algrenewcommand\algorithmicforall{\textbf{\textcolor{algokeyword}{for all}}}
\algrenewcommand\algorithmicreturn{\textbf{\textcolor{algokeyword}{return}}}
\algnewcommand\algorithmicbreak{\textbf{\textcolor{algokeyword}{break}}}

\algrenewcommand{\algorithmiccomment}[1]{\hfill \textcolor{algocomment}{$\triangleright$ \textit{#1}}}

\algrenewcommand\algorithmicrequire{\textbf{Input:}}
\algrenewcommand\algorithmicensure{\textbf{Output:}}

\definecolor{darkspringgreen}{rgb}{0.09, 0.45, 0.27}
\definecolor{denim}{rgb}{0.08, 0.38, 0.74}
\definecolor{darkolivegreen}{rgb}{0.33, 0.42, 0.18}
\definecolor{tangerine}{rgb}{0.95, 0.52, 0.0}
\definecolor{mahogany}{rgb}{0.75, 0.25, 0.0}
\definecolor{coolblack}{rgb}{0.0, 0.18, 0.39}
\definecolor{darkpink}{rgb}{0.91, 0.35, 0.6}
\definecolor{darkblue}{rgb}{0.0, 0.0, 0.67}

\definecolor{seagreen}{rgb}{0.18, 0.55, 0.34}

\definecolor{pred}{rgb}{0.7843, 0.0039, 0.3137} 
\newcommand{\js}[1]{{\color{pred}{#1}}} %

\definecolor{darkpink}{rgb}{0.88, 0.28, 0.54}
\definecolor{forestgreen}{rgb}{0.0, 0.27, 0.13}
\definecolor{amber}{rgb}{1.0, 0.49, 0.0}

\newcommand{\inum}[1]{(\textit{#1})\xspace}

\newcommand{\sects}[1]{{§#1}\xspace} %
\newcommand{\sect}[1]{{§#1}\xspace} %
\newcommand{\head}[1]{{\noindent\textbf{#1.}\xspace}} %
\newcommand{\fig}[1]{{Fig.~#1}\xspace} %

\newcommand\isf{\textsf{Ideal-ISF}\xspace}

\newcolumntype{Y}{>{\centering\arraybackslash}X}
\usepackage{tikz}

\usepackage[compact]{titlesec}
\titlespacing\section{0pt}{3pt plus 2pt minus 2pt}{3pt plus 2pt minus 2pt}
\titlespacing\subsection{0pt}{3pt plus 2pt minus 2pt}{3pt plus 2pt minus 2pt}
\titlespacing\subsubsection{0pt}{3pt plus 2pt minus 2pt}{3pt plus 2pt minus 2pt}

\makeatletter
\g@addto@macro{\normalsize}{%
  \setlength{\abovedisplayskip}{4pt plus 0.5pt minus 1pt}
  \setlength{\belowdisplayskip}{4pt plus 0.5pt minus 1pt}
  \setlength{\abovedisplayshortskip}{0pt}
  \setlength{\belowdisplayshortskip}{0pt}
  \setlength{\intextsep}{3pt plus 1pt minus 1pt}
  \setlength{\textfloatsep}{7pt plus 1pt minus 1pt}
  \setlength{\skip\footins}{4pt plus 1pt minus 1pt}}
\makeatother

\DeclareRobustCommand\wcirc[1]{\tikz[baseline=(char.base)]{           \node[shape=circle,draw,inner sep=0pt,fill=white, text=black] (char) {#1};}}

\usepackage{blindtext,graphicx}
\usepackage[absolute]{textpos}
\usepackage{datetime}

\definecolor{seagreen}{rgb}{0.18, 0.55, 0.34}
\definecolor{ballblue}{rgb}{0.13, 0.67, 0.8}

\definecolor{darkgreen}{rgb}{0.0, 0.44, 0.34}

\sloppypar

\newcommand\hm[1]{{\color{violet}{#1}}}

\definecolor{dollarbill}{rgb}{0.52, 0.73, 0.4}

\renewcommand{\js}[1]{{\color{pred}{#1}}} %

\renewcommand\hm[1]{{\color{denim}{#1}}}

\usetikzlibrary{patterns}
\usepackage[precision=2, unit=mm]{lengthconvert}

\usepackage[colorinlistoftodos,prependcaption,textsize=scriptsize,
textwidth=2cm
]{todonotes}
\definecolor{cyan(process)}{rgb}{0.0, 0.62, 0.82}

\renewcommand{\js}[1]{{\color{pred}{#1}}} %
\renewcommand\hm[1]{{\color{olive}{#1}}}

\newcommand\new[1]{{\color{blue}{#1}}}

\newcommand\proposal{SAGe\xspace}

\renewcommand\new[1]{{\color{black}{#1}}}

\renewcommand{\js}[1]{{\color{black}{#1}}} %
\renewcommand\hm[1]{{\color{black}{#1}}}
\definecolor{cadmiumgreen}{rgb}{0.0, 0.50, 0.29}

\newcommand\revref[1]{\hyperref[rev:#1]{#1}}

\definecolor{dollarbill}{rgb}{0.52, 0.73, 0.4}
\definecolor{olive}{rgb}{0.5, 0.5, 0.0}
\definecolor{green(munsell)}{rgb}{0.0, 0.66, 0.47}
\definecolor{green(ryb)}{rgb}{0.4, 0.69, 0.2}
\definecolor{kellygreen}{rgb}{0.3, 0.73, 0.09}

\definecolor{acolor}{rgb}{0.0, 0.5, 1.0}
\definecolor{bcolor}{rgb}{0.54, 0.17, 0.89}
\definecolor{ccolor}{rgb}{0.4, 0.69, 0.2}
\definecolor{dcolor}{rgb}{0.92, 0.41, 0.12}
\definecolor{ecolor}{rgb}{0.6, 0.0, 0.156}

\definecolor{raspberry}{rgb}{0.89, 0.04, 0.36}

\definecolor{awesome}{rgb}{1.0, 0.13, 0.32}
\definecolor{cardinal}{rgb}{0.77, 0.12, 0.23}
\definecolor{cadet}{rgb}{0.33, 0.41, 0.47}
\definecolor{celadon}{rgb}{0.67, 0.88, 0.69}
\definecolor{persianblue}{rgb}{0.11, 0.22, 0.73}
\definecolor{ultramarine}{rgb}{0.07, 0.04, 0.56}
\definecolor{warmblack}{rgb}{0.0, 0.3, 0.3}

\newcommand\micro[1]{{\color{persianblue}{#1}}}
\newcommand\omcm[1]{{\color{orange}{#1}}}

\renewcommand\micro[1]{{\color{black}{#1}}}
\renewcommand\omcm[1]{{\color{black}{#1}}}

\newcommand\nh[1]{{\color{black}{#1}}}
\definecolor{tealish}{RGB}{0,187,161}
\definecolor{deepred}{RGB}{192,0,0}

\newcommand{\circg}[1]{\tikz[baseline=(char.base)]{
           \node[shape=circle,draw=none,inner sep=0.01pt,fill=tealish!100, text=white] (char) {#1};}}

\newcommand{\circr}[1]{\tikz[baseline=(char.base)]{
           \node[shape=circle,draw=none,inner sep=0.01pt,fill=deepred!100, text=white] (char) {#1};}}

\definecolor{lavenderrr}{rgb}{0.71, 0.49, 0.86}

\newcommand{\sgprop}[1]{
\colorbox{lavenderrr!25}{\textit{\textbf{(Property {#1})}}}}

\definecolor{burgundy}{rgb}{0.5, 0.0, 0.13}

\hypersetup{%
    citecolor=blue,%
    filecolor=blue,%
    linkcolor=blue,%
    urlcolor=blue}

\definecolor{dogwoodrose}{rgb}{0.84, 0.09, 0.41}

\newcommand\omcr[1]{{\color{black}{#1}}}

\definecolor{turquoise}{rgb}{0.19, 0.84, 0.78}
\definecolor{brightturquoise}{rgb}{0.03, 0.91, 0.87}
	\definecolor{cinnamon}{rgb}{0.82, 0.41, 0.12}
    
\newcommand\oii[1]{{\color{black}{#1}}}
\definecolor{azure}{rgb}{0.0, 0.5, 1.0}
\newcommand\oiii[1]{{\color{black}{#1}}}
\definecolor{emerald}{rgb}{0.20, 0.65, 0.38}

\newcommand\oiv[1]{{\color{black}{#1}}}
\definecolor{amethyst}{rgb}{0.85, 0.57, 0.0}
\definecolor{carminered}{rgb}{1.0, 0.0, 0.22}
\definecolor{harvestgold}{rgb}{0.85, 0.57, 0.0}
\definecolor{electricviolet}{rgb}{0.56, 0.0, 1.0}
\definecolor{ochre}{rgb}{0.8, 0.47, 0.13}

\newcommand\ov[1]{{\color{black}{#1}}}
\definecolor{cornflowerblue}{rgb}{0.39, 0.58, 0.93}
\definecolor{darkturquoise}{rgb}{0.0, 0.7, 0.7}
\newcommand\ovi[1]{{\color{black}{#1}}}
\newcommand{\hpcayear}{2026}

\begin{document}

\newcommand{\hpcasubmissionnumber}{2291}
\title{\Large \proposal: A Lightweight Algorithm-Architecture Co-Design for Mitigating\\ 
the Data Preparation Bottleneck in Large-Scale Genome Sequence Analysis}

\def\hpcacameraready{} %
\newcommand{\hpcapubid}{0000--0000/00\$00.00}

\newcommand\hpcaauthors{
Nika Mansouri Ghiasi$^1$ \hspace{0.5em} Talu Güloglu$^1$ \hspace{0.5em} Harun Mustafa$^1$ \hspace{0.5em} Can Firtina$^{1,2}$ \vspace{0em} \\
Konstantina Koliogeorgi$^1$ \hspace{0.5em} Konstantinos Kanellopoulos$^1$ \hspace{0.5em} Haiyu Mao$^3$ \hspace{0.5em} Rakesh Nadig$^1$\vspace{0em}\\
Mohammad Sadrosadati$^1$ \hspace{0.5em} Jisung Park$^4$ \hspace{0.5em} Onur Mutlu$^1$\\\\
}

\newcommand\hpcaaffiliation{\omcr{$^1$ETH Zürich \hspace{0.5em} $^2$University of Maryland \hspace{0.5em} $^3$King's College London \hspace{0.5em} $^4$POSTECH
}}

\author{
\hpcaauthors{}
\hpcaaffiliation{}
}

\fancypagestyle{camerareadyfirstpage}{%
  \fancyhead{}
  \renewcommand{\headrulewidth}{0pt}
  \fancyhead[C]{
    \ifdefined\aeopen
    \parbox[][12mm][t]{13.5cm}{\hpcayear{} IEEE International Symposium on High-Performance Computer Architecture (HPCA)}
    \else
      \ifdefined\aereviewed
      \parbox[][12mm][t]{13.5cm}{\hpcayear{} IEEE International Symposium on High-Performance Computer Architecture (HPCA)}
      \else
      \ifdefined\aereproduced
      \parbox[][12mm][t]{13.5cm}{\hpcayear{} IEEE International Symposium on High-Performance Computer Architecture (HPCA)}
      \else
      \parbox[][0mm][t]{13.5cm}{\hpcayear{} IEEE International Symposium on High-Performance Computer Architecture (HPCA)}    \fi 
    \fi 
    \fi 
    \ifdefined\aeopen 
      \includegraphics[width=12mm,height=12mm]{ae-badges/open-research-objects.pdf}
    \fi 
    \ifdefined\aereviewed
      \includegraphics[width=12mm,height=12mm]{ae-badges/research-objects-reviewed.pdf}
    \fi 
    \ifdefined\aereproduced
      \includegraphics[width=12mm,height=12mm]{ae-badges/results-reproduced.pdf}
    \fi
  }
  \fancyfoot[C]{}
}
\fancyhead{}
\renewcommand{\headrulewidth}{0pt}

\maketitle

\ifdefined\hpcacameraready 
  \thispagestyle{camerareadyfirstpage}
  \pagestyle{empty}
\else
  \thispagestyle{plain}
  \pagestyle{plain}
\fi

\newcommand{\hpcaheight}{0mm}
\ifdefined\eaopen
\renewcommand{\hpcaheight}{12mm}
\fi

\setcounter{page}{1}
\begin{abstract}

  \nh{Genome sequence analysis, which \oii{examines} the DNA sequences of organisms, drives advances in many critical medical and biotechnological fields.}
Given \nh{its importance and} the exponentially growing volumes of genomic \nh{sequence} data, there are extensive efforts to accelerate genome sequence analysis. \nh{In this work,} we demonstrate a major bottleneck that greatly limits and diminishes the benefits of state-of-the-art genome sequence analysis accelerators: the \emph{data preparation bottleneck}, where genomic \nh{sequence} data is stored in compressed form and needs to be \omcr{first} decompressed and formatted before an accelerator can operate on it. To mitigate this bottleneck, we propose \textbf{\proposal}, an algorithm-architecture co-design for highly-compressed \textbf{\underline{s}}torage and high-performance \textbf{\underline{a}}ccess of large-scale \textbf{\underline{ge}}nomic \nh{sequence} data. \nh{The key challenge} is to improve data preparation performance while maintaining high compression ratios (comparable to genomic-specific compression algorithms) at low hardware cost. \nh{We address this challenge} by leveraging key properties of genomic datasets to co-design \inum{i}~a lossless (de)compression algorithm, \inum{ii}~hardware \nh{that decompresses data with lightweight operations and efficient streaming accesses}, \inum{iii}~storage data layout, and \inum{iv}~interface commands to access data. \nh{\proposal is highly versatile, as it supports datasets from different sequencing technologies and species}.
\omcr{Due} to its lightweight design, \proposal 
can be seamlessly integrated with a broad range of \oii{hardware accelerators for} genome sequence analysis to mitigate their data preparation bottlenecks.  
Our results demonstrate that \proposal improves the average end-to-end performance and energy efficiency of two state-of-the-art genome sequence analysis accelerators by 3.0$\times$--32.1$\times$ and \nh{13.0$\times$--34.0$\times$}, respectively, compared to when the accelerators rely on state-of-the-art \omcr{software and hardware} decompression tools.

\end{abstract}

 \section{Introduction}
\label{sec:intro}
Genome sequence analysis plays an important role in many fields, such as personalized medicine~\cite{clark2019diagnosis,farnaes2018rapid,sweeney2021rapid,alkan2009personalized,flores2013p4,ginsburg2009genomic,chin2011cancer,Ashley2016}, tracing outbreaks of communicable diseases~\cite{bloom2021massively,yelagandula2021multiplexed,le2013selected,nikolayevskyy2016whole,qiu2015whole,gilchrist2015whole}, \micro{ensuring food safety through pathogen monitoring~\cite{e002244,TONG2021130}}, \omcr{agriculture~\cite{prasad2021soil,Mascher2024,Schreiber2024}, scientific discovery~\cite{urbanek2018degradation,edgar2022petabase,paoli2022biosynthetic}, biodiversity conservation~\cite{Hogg2024,lewin2018earth}}, \oii{evolutionary biology~\cite{ellegren2014genome,Prado-Martinez2013,Prohaska2019}, and antimicrobial resistance surveillance~\cite{danko2021global,didelot2012transforming,marini2022towards}}.
To analyze genomic information computationally, an organism's DNA sample\footnote{\omcr{Even if an organism's genome is RNA-based, it is typically converted to DNA before sequencing~\cite{Houldcroft2017,Jansz2024viral}.}} undergoes a process called \emph{sequencing}, which converts the information from DNA molecules to digital data. Current sequencing technologies \emph{cannot} sequence long DNA molecules end-to-end. Instead, state-of-the-art sequencers generate randomly- and redundantly\hm{-}sampled smaller and inexact DNA fragments, called \emph{reads}. Sets of genomic reads (called \emph{read sets}) are then used in genome analysis. 
The importance of genome sequence analysis, along with the rapid improvement in sequencing \nh{technologies} (i.e., reduced costs and increased throughput~\cite{berger2023navigating}), has led to the rapidly increasing adoption of genome analysis in recent years \cite{clark2019diagnosis,farnaes2018rapid,sweeney2021rapid,ginsburg2009genomic,chin2011cancer,Ashley2016,bloom2021massively,gilchrist2015whole} and continuous exponential growth in genomic data generation~\cite{katz2021sra,ena2022,srastats,enastats} (far outpacing Moore's \omcr{L}aw~\cite{stephens2015big}).

It is common practice to store genomic \nh{sequence} data in compressed forms\omcr{~\cite{berger2023navigating,zhu2013high,Deorowicz2013,giancarlo2013compressive,Betschart2025}} because storing uncompressed \omcr{genomic} data is impractical \omcr{due to its massive size}. In fact, due to the importance of storing genomic \nh{sequence} data in a space-efficient manner, there exist many compression techniques (e.g.,\omcr{~\cite{chandak2018spring,chandak2017compression,roguski2018fastore,cogo2021genodedup,kokot2022colord,Meng2023,dufort2021renano,kowalski2019pgrc,dufort2020enano,dragenora,yang2025gpufastqlz,chen2023efficient,hach2012scalce,roguski2014dsrc2,Deorowicz2020,lan2021genozip,alyami2019lfastqc}}) specialized for genomic \nh{sequence} data to achieve significantly higher compression ratios than state-of-the-art general-purpose compress\omcr{ion methods (e.g.,~\cite{collet2018zstandard,pavlov20167,Brotli,Katz1991US5051745A,goyal2021dzip,goyal2018deepzip,chen2024ha,bartik2015lz4,liu2018data,fowers2015scalable,chen2021fpga,angerd2022gbdi,gao2024beezip,karandikar2023cdpu,9499902,abali2020data})}.

\nh{Due to the challenges of analyzing massive volumes of genomic data, and its pivotal importance in many critical fields,} there are extensive efforts to accelerate genome sequence analysis\omcr{~\cite{alser2020accelerating,alser2022molecules,mutlu2023accelerating}}. These efforts focus on alleviating the computational overheads \omcr{of genome sequence analysis} via \emph{algorithmic optimizations} (e.g.,\omcr{~\cite{zhang2000greedy,slater2005automated,li2018minimap2,myers1999fast,marco2021fast,marcosola2023optimal,grootkoerkamp2024apa2,xin2013accelerating,xin2015shifted}}) or \emph{hardware accelerators} (e.g.,\omcr{~\cite{doblas2025smx,kim2025nmp,mutlu2023accelerating,alser2022molecules,lou2020helix,lou2018brawl,shahroodi2023swordfish,saavedra2020mining,markus2020benchmarking,subramaniyan2021accelerated,huangfu2018radar,khatamifard2021genvom, cali2020genasm, gupta2019rapid,li2021pim,angizi2019aligns,zokaee2018aligner,turakhia2018darwin, fujiki2018genax, madhavan2014race,cheng2018bitmapper2,houtgast2018hardware,houtgast2017efficient, zeni2020logan,ahmed2019gasal2,nishimura2017accelerating,de2016cudalign,liu2015gswabe,liu2013cudasw++,liu2009cudasw++,liu2010cudasw++,wilton2015arioc,goyal2017ultra,chen2016spark,chen2014accelerating,chen2021high,fujiki2020seedex, banerjee2018asap,fei2018fpgasw,waidyasooriya2015hardware,chen2015novel,rucci2018swifold,haghi2021fpga,li2021pipebsw,ham2020genesis,ham2021accelerating,wu2019fpga,kaplan2020bioseal}}), and/or reducing \nh{its} data movement overheads via \emph{near-data processing} (e.g., in main memory\oii{~\cite{wu2021sieve,shahroodi2022krakenonmem,shahroodi2022demeter,dashcam23micro,hanhan2022edam,zou2022biohd,
cali2020genasm, huangfu2018radar, khatamifard2021genvom, gupta2019rapid, li2021pim, angizi2019aligns, zokaee2018aligner,Zhang_2023_alignerD,soysal2025mars,cali2022segram,kim2018grim,kaplan2020bioseal,mao2022genpip}} or storage\omcr{~\cite{mansouri2022genstore,abakus23taco,megis,jun2016storage,kim2025nmp,soysal2025mars,zheng2025storage}}). \omcr{Unfortunately, all these acceleration efforts are limited greatly by the data preparation bottleneck (which no prior work has fully alleviated), as we describe next.}

\nh{\head{Data Preparation Bottleneck in Genome Sequence Analysis} We \oii{empirically} demonstrate that the benefits of prior works on accelerating genome sequence analysis are greatly diminished due to what we call the \emph{data preparation bottleneck}, where compressed genomic \nh{sequence} data needs to be \omcr{first} decompressed and formatted before it can be analyzed. 
For example, \fig{\ref{fig:intro-motivation}} shows the execution timeline of data preparation and genome analysis for a real-world genomic dataset (\sect{\ref{sec:methodology}}) in three different configurations. The evaluated analysis task is read mapping, a fundamental process in genomics (\sect{\ref{sec:background-workflow}}). The configurations are 
\inum{i}~\textbf{Base\omcr{line}:}~a state-of-the-art software analysis tool~\cite{li2018minimap2} with a state-of-the-art software genomic decompressor for data preparation~\cite{chandak2018spring};
\inum{ii}~\textbf{Acc. Analysis:} a state-of-the-art hardware-accelerated analysis tool~\cite{chen2023gem} with the same \omcr{data} preparation \omcr{as Baseline};
\inum{iii}~\textbf{Acc. Analysis w/ Ideal Prep.:} the same accelerated analysis with \emph{ideal} preparation, where preparation is overlapped \omcr{with} analysis. 
For all configurations, preparation and analysis operate in a pipelined manner and in batches (i.e., when decompressing  batch $\#i$, the mapper
analyzes batch $\#i - 1$).\footnote{\nh{Decompressed data batches are directly fed to the analysis stage.}}
We observe that \circg{1}~hardware acceleration of genome analysis can potentially offer substantial performance benefits; \circr{2}~however, as analysis gets faster, data preparation emerges as a critical bottleneck that hinders the full realization of these benefits. Our motivational study (\sect{\ref{sec:motivation}}) further demonstrates this bottleneck across various real-world \omcr{scenarios}.}

\begin{figure}[t]
         \centering
         \includegraphics[width=\columnwidth]{Figures/motivation-brand-new.pdf}
         \caption{\nh{Effect of data preparation (i.e., decompressing and formatting genomic sequence data before analysis) on genome analysis performance.}}
         \label{fig:intro-motivation}
\end{figure}

\head{Requirements for Mitigating the Data Preparation Bottleneck} To fully benefit from the extensive efforts on accelerating genome sequence analysis, without resorting to storing massive volumes of genomic \nh{sequence} data uncompressed, there is a critical need for a design that effectively mitigates the data preparation bottleneck while meeting three key requirements. First, it should achieve \emph{high performance and energy efficiency}. Second, it should maintain \emph{high compression ratios}, comparable to state-of-the-art genomics-specific compression techniques.  Third, it must be \emph{lightweight} for seamless integration with a broad range of genome analysis systems, including general-purpose \omcr{systems}, specialized accelerators, portable genomics systems, and near-data processing (NDP) architectures.

\head{Challenges} Meeting all three requirements simultaneously is challenging. Some prior works (e.g.,\oii{~\cite{guo2013gpu,qiao2019fpga,zhao2017streaming,jiang2021exma,arram2015fpga,wang2018accelerating,chen2023efficient,leavline2013hardware,lim2025bancroft}}) accelerate certain computation kernels that are widely used in many state-of-the-art genomic compressors (e.g.,\oii{~\cite{lan2021genozip,Meng2023,kowalski2019pgrc,chandak2017compression,chen2023efficient}}). Despite their benefits, these approaches require expensive computational units, large buffers, or large DRAM bandwidth or capacity (e.g., due to many random accesses to large data structures for matching patterns). Therefore, none of these approaches is suitable for seamless integration with genome analysis accelerators, particularly when targeting integration into resource-constrained environments, such as various NDP accelerators (e.g.,\oii{~\cite{wu2021sieve,shahroodi2022krakenonmem,shahroodi2022demeter,dashcam23micro,hanhan2022edam,zou2022biohd, cali2020genasm, huangfu2018radar, khatamifard2021genvom, gupta2019rapid, li2021pim, angizi2019aligns, zokaee2018aligner,Zhang_2023_alignerD,mansouri2022genstore,abakus23taco,megis,jun2016storage,soysal2025mars,kim2025nmp,zheng2025storage,cali2022segram,kim2018grim,kaplan2020bioseal,mao2022genpip,angizi2020pim}}) or portable genomics devices (e.g.,\oii{~\cite{MinIONMk1CO,palatnick2020igenomics,Ballard2018,Oehler2023,watsa2020portable,Quick2016,wang2021nanopore}}), which play critical roles in genomics.

\textbf{Our goal} in this work is to improve the performance and energy efficiency of genome sequence analysis by mitigating the data preparation bottleneck, while maintaining high compression ratios and ensuring a lightweight design. To this end, we propose \textbf{\proposal}, an algorithm-architecture co-design for highly-compressed \textbf{\underline{s}}torage and high-performance \textbf{\underline{a}}ccess of large-scale \textbf{\underline{ge}}nomic \nh{sequence} data. \nh{\proposal is based on the \textbf{key insight} that the information encoded in genomic data follows specific trends, shaped by factors such as sequencing technology (e.g., error rates and read lengths) and common genetic phenomena (e.g., typical \omcr{spatial distributions of} genetic variations \omcr{within genomes}). By carefully exploiting these \oii{characteristics to} synergistically co-design algorithms and hardware, \proposal achieves high compression ratios, comparable to state-of-the-art genomic compressors, while enabling \oii{low} decompression \oii{latencies}, using only lightweight hardware and efficient streaming accesses.}

\head{Key Mechanism} Leveraging our rigorous analysis of genomic datasets' properties, we propose an efficient and synergistic co-design that comprises four aspects. First, \proposal{} stores information \oii{about} reads in a genomic read set in array structures that, during decompression, can be interpreted by efficient accesses and lightweight \nh{operations}. To maintain high compression ratios using \emph{only} these lightweight structures, \proposal's lossless compression algorithm \nh{optimizes} the encoding of these structures based on the properties of \emph{each genomic dataset} \nh{to accommodate datasets sequenced with different sequencing technologies and from different species}. \omcr{Examples of these dataset properties include sequencing error rates, read lengths, and spatial distributions of genetic variations.} Second, we design lightweight hardware units to efficiently decompress \proposal's data structures. \proposal's low-cost and lightweight design enables seamless integration with a wide range of genome sequence analysis accelerators, without competing for resources (e.g., memory bandwidth and capacity) required by the genome sequence analysis tasks. Third, we design an efficient data layout to leverage the storage system's full bandwidth when accessing compressed genomic data. Fourth, we design specialized interface commands that are exposed to genomics applications to access data and communicate with \proposal's hardware to decompress the data to the desired format.

\head{Key Results} 
\omcr{We evaluate the end-to-end performance of \oii{various genome analysis} applications, where execution includes \emph{both} data preparation and genome analysis.} 
For data preparation, we compare \proposal to the following:
\inum{i}~pigz~\cite{adler2015pigz}, a widely-used general-purpose compressor;\footnote{\nh{We exclude other general‐purpose compressors from our performance analysis since, as detailed in \sect{\ref{sec:background-compression}}, they deliver substantially lower compression ratios than genomic compressors, and thus do not align with the field’s move toward genomic compressors~\cite{hernaez2019genomic,dragenora}. We include pigz\omcr{~\cite{adler2015pigz}}, however, since it remains a common comparison baseline in genomic compression studies.}}
\inum{ii}~Spring~\cite{chandak2018spring} and NanoSpring~\cite{Meng2023}, state-of-the-art genomics-specific compressors for short and long reads, respectively;
and \inum{iii}~\nh{hardware-accelerated} Spring and NanoSpring \oii{(see \sect{\ref{sec:methodology}} for details)}.
For genome analysis, we evaluate: \nh{\inum{i}~GEM~\cite{chen2023gem}, a state-of-the-art genome analysis accelerator, and \inum{ii}~GenStore~\cite{mansouri2022genstore}, a state-of-the-art NDP genome analysis accelerator inside the resource-constrained environment of an SSD.} 
We show that \nh{when \omcr{each configuration is} integrated with GEM,} \proposal improves performance by an average of 12.3$\times$, 3.9$\times$, and 3.0$\times$, and energy efficiency by \nh{34.0$\times$, 16.9$\times$, and 13.0$\times$} over pigz, (Nano)Spring, and hardware-accelerated (Nano)Spring, \omcr{respectively}.
\nh{When \omcr{each configuration is} integrated with GenStore,} \proposal provides 32.1$\times$, 10.4$\times$, and 7.8$\times$ average speedup over pigz, (Nano)Spring, and hardware-accelerated (Nano)Spring, respectively, \oii{at a} very low area cost of 0.7\% of the three cores~\cite{cortexr4} in an SSD controller~\cite{samsung860pro}.
\proposal provides these benefits while 
achieving significantly larger compression ratios than pigz (2.9$\times$ larger, on average) and comparable ratios to (Nano)Spring (with \omcr{an} average reduction of only 4.6\%).

This work makes the following \textbf{key contributions}:
\begin{itemize}[leftmargin=*, noitemsep, topsep=0pt]
    \item We demonstrate that the \emph{data preparation bottleneck} can greatly \oii{limit} the potential \oii{performance and efficiency} benefits of genome sequence analysis hardware accelerators.
    \item We propose \textbf{\proposal}, an algorithm-system co-design for highly-compressed storage and high-performance access of genomic data, to mitigate the data preparation bottleneck, \omcr{an important bottleneck that is not assessed or handled by prior works on genome analysis acceleration.}
    \item We leverage properties of genomic data to co-design \proposal's algorithm and architecture, such that highly-compressed data can be \oii{efficiently} interpreted by lightweight hardware and rapidly prepared for analysis. 
    \item \proposal's key contribution lies in the synergistic and genomic\omcr{-}data-aware co-design of algorithms, lightweight hardware, data layout, and a specialized software interface. This synergistic co-design is the key enabler of \proposal's lightweight, high-performance, and energy-efficient design.
    \omcr{\item We demonstrate that \proposal significantly improves the end-to-end performance and energy efficiency of state-of-the-art genome sequence analysis \oii{hardware} accelerators, compared to when the accelerators rely on state-of-the-art software and hardware decompression tools.}
    \end{itemize}

\section{Background}
\label{sec:background}

\begin{figure}[b]
  \centering
  \includegraphics[width=\columnwidth]{Figures/background-workflow.pdf}
  \caption{Overview of a typical genomic workflow.}
  \label{fig:workflow-overview} 
\end{figure}

\subsection{Genomic Workflow}
\label{sec:background-workflow}
Given a genomic sample, a typical workflow consists of three key steps\oii{~\cite{alser2020technology,alser2020accelerating,berger2023navigating,alser2022molecules}}, as shown in \fig{\ref{fig:workflow-overview}}. The first step, \emph{sequencing} (\circled{1}), converts DNA into digital signals. Sequencers cannot read many organisms' entire genomes as complete, error-free sequences. Instead, they produce many shorter \nh{overlapping} sequences called \emph{reads} that are randomly sampled from different locations in the genome. 
The average number of reads per location is called the \emph{sequencing depth}. Deeper sequencing provides more information per location, which allows later analysis steps to better distinguish between biological signals and sequencing errors. Sequencers are distinguished by the characteristics of their reads.  
Examples include sequencers that produce short reads (75 to 300 characters) with high accuracy ($\sim$99.9\%)\oiii{~\cite{stoler2021sequencing,goodwin2016coming,davis2021sequencerr,sereika2022oxford}} and long reads (\oiii{typically 500} to 25k characters\oiii{, and up to $\sim$2.2M characters~\cite{jain2018nanopore,payne2018bulkvis,amarasinghe2020opportunities}}) with intermediate accuracy ($\sim$99\%)\oiii{~\cite{hon2020highly,ni2023benchmarking,wenger2019accurate, amarasinghe2020opportunities, sereika2022oxford}}.

Second, \emph{basecalling} (\circled{2})\oii{~\cite{cock2009sanger,alser2022molecules,singh2024rubicon,xu2021fast,lou2020helix,shahroodi2023swordfish,Samarakoon2023accelerated,cavlak2022targetcall}}, converts the sequencer's raw signals to strings. 
Each DNA read is converted to: \inum{i}~the DNA encoded using an alphabet representing the nucleotides (also called base pairs) A, C, G, T (along \ov{with} N to represent unknowns), \inum{ii}~a quality score~\cite{ewing1998base} for each nucleotide, encoding its probability of being incorrect using an ASCII character, and \inum{iii}~a header. 
The strings from all reads in a sample \oii{(i.e., a read set)} are then written to a file.\footnote{This format is FASTQ~\cite{cock2009sanger}, the most common read set format~\cite{illuminafastq}.} The file is then typically stored compressed  (\sect{\ref{sec:background-compression}}) for future analysis.

The third step is \emph{genome sequence analysis} (\circled{3}) on a collection of read sets. 
\nh{Since read sets are typically stored compressed\omcr{~\cite{berger2023navigating,zhu2013high,Deorowicz2013,giancarlo2013compressive,Betschart2025}}, they need to be prepared (i.e., decompressed and formatted) before analysis.}
A typical analysis workflow quantifies mismatches between the reads and a reference genome. This typically starts with the computationally-expensive \emph{read mapping} process\oii{~\cite{alser2020accelerating, alser2020technology,alser2022molecules,li2018minimap2,langmead2012fast,kim2019graph,li2013aligningsequencereadsclone}}, which finds the potential matching locations of reads in the reference genome. The result of read mapping can then be fed to various downstream analysis tasks (e.g., variant calling\oii{~\cite{alser2022molecules,Olson2023,Poplin2018,li2008mapping,Yu2015,Park2025}}, genome assembly\oii{~\cite{alser2022molecules,firtina2020apollo,turakhia2018darwin,cali2017nanopore,li2016minimap,Haghshenas2020,Zimin2017hybrid}}, \nh{and taxonomic classification\oii{~\cite{wu2021sieve,wood2019improved,wood2014kraken,truong2015metaphlan2,kim2016centrifuge,song2024centrifuger,ounit2015clark,piro2016dudes,Fan2021,piro2020ganon,Marcelino2020,milanese2019microbial,lapierre2020metalign,pockrandt2022metagenomic,Dilthey2019,lemane2023kmindex,shen2022kmcp,sun2021challenges,lu2017bracken,koslicki2016metapalette,dimopoulos2022haystac,meyer2021critical}}}). 
Analysis workflows \oii{typically} access all \oii{base pairs}, but only a small fraction of the quality scores. This is because read mapping, while using all \oii{base pairs}, typically ignores quality scores\oii{~\cite{li2018minimap2,vasimuddin2019efficient,Meng2023,kolmogorov2019assembly,Cheng2021}}. The subsequent steps (e.g., variant calling) only need quality scores from the positions surrounding mismatches\oii{~\cite{Poplin2018,li2008mapping,Yu2015,Park2025}} to distinguish true biological variation from sequencing errors.

The analysis step often begins with \emph{existing, already sequenced and basecalled read sets} because, in many cases, a single read set needs to be analyzed \emph{many times} and at \emph{different times}\oii{~\cite{aganezov2022complete,berger2023navigating,Siren2024,vaddadi2023minimizing,khayat2021hidden}}. 
For example, some applications (e.g.,~measuring population genetic diversity\oii{~\cite{kostlbacher2021pangenomics,vandorp2020emergence,Logsdon2025,Zheng2017alignment,aganezov2022complete}}) require analyzing a read set with many reference genomes at different times.
Other applications require repeating the analysis many times (e.g.,~with updated or personalized references\oii{~\cite{chen2024improved,Siren2024,vaddadi2023minimizing,aganezov2022complete,berger2023navigating,rhie2023complete,nurk2022complete,kim2024airlift}}) to improve accuracy. 
Due to the criticality of the analysis step and the challenges of analyzing large amounts of sequencing reads on conventional systems, a large body of works (e.g.,\oii{~\cite{mutlu2023accelerating,alser2020accelerating,alser2022molecules,lou2020helix,lou2018brawl,shahroodi2023swordfish,saavedra2020mining,markus2020benchmarking,subramaniyan2021accelerated,huangfu2018radar,khatamifard2021genvom, cali2020genasm, gupta2019rapid,li2021pim,angizi2019aligns,zokaee2018aligner,turakhia2018darwin, fujiki2018genax, madhavan2014race,cheng2018bitmapper2,houtgast2018hardware,houtgast2017efficient, zeni2020logan,ahmed2019gasal2,nishimura2017accelerating,de2016cudalign,liu2015gswabe,liu2013cudasw++,liu2009cudasw++,liu2010cudasw++,wilton2015arioc,goyal2017ultra,chen2016spark,chen2014accelerating,chen2021high,fujiki2020seedex, banerjee2018asap,fei2018fpgasw,waidyasooriya2015hardware,chen2015novel,rucci2018swifold,haghi2021fpga,li2021pipebsw,ham2020genesis,ham2021accelerating,wu2019fpga,cali2022segram,kim2018grim,kim2025nmp,doblas2025smx,zhang2000greedy,slater2005automated,%
jia2011metabing,kobus2021metacache,wang2023gpmeta,kobus2017accelerating,su2013gpumetastorms,%
Yano2014,Gamaarachchi2020_f5c,%
zhang2023genomix,cervi2022metagenomic,shih_haru_2023,%
liyanage2023efficient,%
wu2021sieve,shahroodi2022krakenonmem,shahroodi2022demeter,dashcam23micro,hanhan2022edam,zou2022biohd,%
Zhang_2023_alignerD, mansouri2022genstore,abakus23taco,megis,jun2016storage,%
singh2024rubicon,xin2013accelerating,firtina2024aphmm,xin2015shifted,alser2017gatekeeper,alser2019shouji,alser2017magnet,alser2020sneakysnake,bingol2021gatekeeper,xin2016optimal,mao2022genpip,kaplan2020bioseal,angizi2020pim}}) accelerate genome sequence analysis tasks, particularly read mapping (e.g.,\oii{~\cite{alser2020accelerating, huangfu2018radar, cali2020genasm, turakhia2018darwin, fujiki2018genax, fujiki2020seedex, banerjee2018asap, khatamifard2021genvom, gupta2019rapid, li2021pim, angizi2019aligns, zokaee2018aligner, madhavan2014race, cheng2018bitmapper2, houtgast2018hardware,houtgast2017efficient, goyal2017ultra, chen2016spark, chen2014accelerating, chen2021high, zeni2020logan, ahmed2019gasal2, nishimura2017accelerating, de2016cudalign, liu2015gswabe, liu2013cudasw++, wilton2015arioc, fei2018fpgasw, waidyasooriya2015hardware, chen2015novel, rucci2018swifold, haghi2021fpga, li2021pipebsw, ham2020genesis, ham2021accelerating, wu2019fpga,doblas2025smx,kim2018grim,alser2020sneakysnake,bingol2021gatekeeper,xin2016optimal,mao2022genpip,mansouri2022genstore,xin2015shifted,alser2017gatekeeper,alser2019shouji,alser2017magnet,Zhang_2023_alignerD,cali2022segram,mutlu2023accelerating,alser2022molecules,subramaniyan2021accelerated,xin2013accelerating,kaplan2020bioseal,angizi2020pim}}), which is one of the most critical bottlenecks in many analysis applications~\cite{berger2023navigating}.

\subsection{Storing Genomic Sequence Data}
\label{sec:background-compression}

\nh{\head{Criticality of Genomic Sequence Data}} Driven by continuing exponential drops in sequencing costs~\cite{nhgri,pedro2021integration} and the pivotal importance of genome sequence analysis, 
genomic \nh{sequence} data volumes 
in public~\cite{katz2021sra,srastats,enastats} and private~\cite{Bick2024,Li2023whole} repositories 
are growing by an \emph{order of magnitude} every few years, approaching and expected to exceed the data volumes generated by various major \oii{internet} media platforms~\cite{stephens2015big,srastats,katz2021sra,enastats}.
While \oii{genome analysis also works on} other data types (e.g., assembled genomes\oii{~\cite{Garrison2024,Armstrong2020,Minkin2020}}, epigenetic markers\oii{~\cite{Chen2025,Liu2025,Sigurpalsdottir2024,Li2011}}, and 3D chromosomal contact maps\oii{~\cite{Forcato2017,Pal2019,lieberman2009comprehensive}}), 
analyzing sequence data is one of the most critical processes in genomics, with sequence data constituting the largest fraction of data stored and analyzed~\cite{ncbi2025}. For example, sequence data is the \emph{largest} class of data maintained at institutions like the National Center for Biotechnology Information and the European Molecular Biology Laboratory~\cite{thakur2023embl,ncbi2025}. 
Note that many other genomic data types \oii{(e.g., assembled genomes, epigenetic markers, and 3D chromosomal contact maps)} are also derived from sequence data, so it is critical to store this \oii{sequence} data to ensure both reproducibility and to enable reanalysis with future improved workflows. 
For example, while some tasks (e.g., variant calling) can start with already mapped reads, best-practice guidelines dictate that reads should be remapped during the course of a study to ensure the use of appropriate (updated\ov{~\cite{chen2024improved,aganezov2022complete,berger2023navigating,Siren2024,rhie2023complete,nurk2022complete}} or personalized\ov{~\cite{Siren2024,vaddadi2023minimizing}}) reference genomes and mapping parameters.
This is reflected in the fact that, as of October 2025, 75.9\% \ov{(18 peta DNA bases)} of publicly-deposited whole-genome sequencing read sets~\cite{katz2021sra,ncbi2025} are in unmapped, i.e., FASTQ format~\cite{cock2009sanger}\ov{, and their overall size has been growing at a rate of 42.5\% on average per year over the last decade}.

\head{Genomic-Specific Compression} Due to the importance of storing genomic data space-efficiently, there exist many compression techniques (e.g.,\omcr{~\cite{chandak2018spring,chandak2017compression,kokot2022colord,Meng2023,roguski2018fastore,dufort2021renano,kowalski2019pgrc,dufort2020enano,dragenora,yang2025gpufastqlz,chen2023efficient,hach2012scalce,roguski2014dsrc2,Deorowicz2020,lan2021genozip,alyami2019lfastqc,cogo2021genodedup}}) specialized for genomic data. \oii{These techniques attain} higher compression ratios (e.g.,~typically from $\sim$2 to $\sim$40~\cite{chandak2018spring,roguski2018fastore,dufort2020enano,dufort2021renano}) than general-purpose compressors (e.g.,~typically from $\sim$2 to $\sim$6~\cite{chandak2018spring}), \oii{because general-purpose approaches fail to leverage longer-range similarities common in DNA data~\cite{hernaez2019genomic,zhu2013high}.} 
This large gap \oii{in compression ratios exists} even with state-of-the-art general-purpose compressors in software (e.g.,~\cite{collet2018zstandard,pavlov20167,Brotli,Katz1991US5051745A,goyal2021dzip,goyal2018deepzip,chen2024ha}) or hardware (e.g.,~\cite{bartik2015lz4,liu2018data,fowers2015scalable,chen2021fpga,angerd2022gbdi,gao2024beezip,karandikar2023cdpu,9499902,abali2020data}), which is the reason for the \oii{large} emphasis in genomics to \oii{use} genomic\oii{-specific} compressors~\cite{hernaez2019genomic,dragenora}.
For example, \oii{hardware-based} Intel QAT~\cite{intelqat} and IBM zEDC~\cite{ibmzedc} \oii{compression accelerators} lead to $2.6 \times$ and $2.7 \times$ lower average compression ratios, respectively, than genomic compressors~\cite{chandak2018spring,Meng2023} on our datasets (\sect{\ref{sec:methodology}}). \oii{Similarly,} state-of-the-art general compression algorithms such as xz~\cite{xz} (a state-of-the-art LZMA-based compressor) and zstd~\cite{collet2018zstandard}, both at their highest levels, \oii{lead to} 2.13$\times$ average (up to 6.7$\times$) worse compression ratios \oii{than genomic compressors~\cite{chandak2018spring,Meng2023} on our datasets}.

\begin{figure}[t]
  \centering
  \includegraphics[width=\columnwidth]{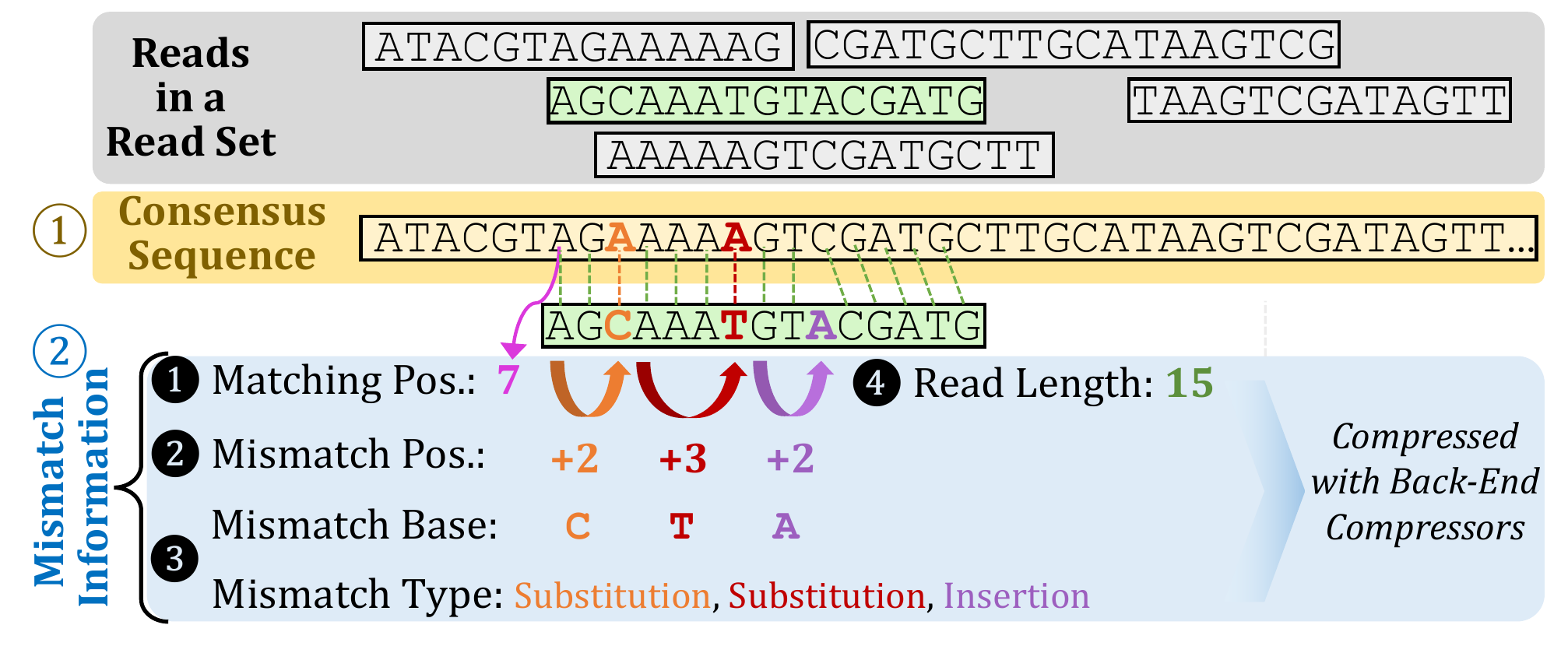}
  \caption{\new{Overview of genomics-specific compression.}}
  \label{fig:genomic-compression} 
\end{figure}

\fig{\ref{fig:genomic-compression}} shows an example of a typical genomic compression technique. 
Genomic compressors typically represent \nh{the DNA bases in} each read set with a \wcirc{1}~\emph{consensus sequence} and \wcirc{2}~the \emph{mismatches} of each read in the read set compared to that consensus~\cite{chandak2017compression,dufort2021renano,kokot2022colord,kowalski2019pgrc}. A consensus sequence is an approximation of the organism's genome, and can either be a user-provided reference~\cite{dufort2021renano} or a de-duplicated string derived from the reads\hm{, representing the most likely character at each location}~\cite{chandak2017compression,kowalski2019pgrc}. Note that it is not sufficient to store only the consensus to represent the read set because the consensus alone does not capture the \oii{variations} in the original reads. Individual reads often contain unique differences, including sequencing errors~\cite{goodwin2016coming,wang2021nanopore} or biological variation~\cite{Cheng2022}.

Given a consensus sequence, a lossless encoding of a read's DNA bases consists of \circled{1}~\oii{the read's} matching position in the consensus \oii{sequence} (many compressors, e.g.,~\cite{chandak2017compression,Meng2023,roguski2018fastore,kowalski2019pgrc}, store matching positions based on their order of appearance in the consensus, which enables space-efficient delta-encoding),
\circled{2}~mismatch positions (delta-encoded), 
\circled{3}~mismatch bases and types (i.e., substitution, insertion, deletion), and
\circled{4}~read length (\oii{especially} for long reads since they have variable lengths). This encoding of the \nh{DNA bases} is then more compressible using general\oii{-purpose} compressors~\cite{chandak2017compression}, which are then used by the state-of-the-art genomic compressors to further compress the mismatch information. 
\nh{Since quality scores do not have the same redundancy patterns as DNA bases, state-of-the-art genomic compressors process them in separate streams and compress them using various techniques (e.g., lossless context models~\cite{dufort2020enano,kokot2022colord,Deorowicz2020,alyami2019lfastqc,cogo2021genodedup} or block sorting~\cite{chandak2018spring}, and lossy quantization~\cite{chandak2017compression}).
Some tools (e.g.,~\cite{Meng2023,vandamme2024tinted,karasikov2020metagraph}) discard quality scores altogether since many recent workflows (e.g.,~\cite{li2018minimap2,wood2019improved,song2024centrifuger,kolmogorov2019assembly,colquhoun2021pandora}), particularly those using accurate reads, do not use quality scores.}

\section{Motivational Analysis}
\label{sec:motivation}

\newcommand\pigz{\texttt{pigz}\xspace}
\newcommand\nspring{\texttt{(N)Spr}\xspace}
\newcommand\nsacc{\texttt{(N)SprAC}\xspace}
\newcommand\ideal{\texttt{Ideal}\xspace}

\nh{We show the impact of \nh{the} data preparation bottleneck in \sect{\ref{sec:motivation-observations}} and discuss challenges of mitigating it in \sect{\ref{sec:motivation-goal}}.}

\subsection{\nh{Data Preparation Bottleneck in Genome Sequence Analysis \oii{Accelerators}}}
\label{sec:motivation-observations}

\nh{As shown in \fig{\ref{fig:intro-motivation}}, when genome sequence analysis is accelerated, data preparation emerges as a critical bottleneck. We perform experimental studies to understand the impact of this bottleneck when analyzing different real-world read sets.}

\head{Methodology} 
\omcr{We evaluate the end-to-end performance of \oii{a genome analysis} application, where execution includes both data preparation and genome analysis.}
For \emph{genome sequence analysis}, we use a state-of-the-art hardware accelerator~\cite{chen2023gem} for read mapping. 
For \emph{data preparation}, \nh{we consider the following configurations to decompress data into the desired uncompressed format:
\inum{i}~\pigz: A parallel version~\cite{adler2015pigz} of gzip, a commonly-used general compressor;
\inum{ii}~\nspring: Spring~\cite{chandak2018spring} and NanoSpring~\cite{Meng2023}, state-of-the-art \oii{software} compressors for short and long reads, respectively; and
\inum{iii}~\ideal: an idealized compressor with zero decompression time.} For our evaluated real-world read sets (\sect{\ref{sec:methodology}}),
Spring and pigz achieve an average compression ratio of 16.9 and 5.4, respectively. 
\nh{We exclude other general-purpose compressors from our performance evaluations since, \nh{as detailed in \sect{\ref{sec:background-compression}}}, they achieve significantly worse compression ratios than genomic compressors, and thus do not align with the \oii{substantial need} to \oii{use} genomic compressors~\cite{hernaez2019genomic,dragenora}.
We do, however, include pigz \oii{(}parallel gzip\oii{)} since it is still widely used as a baseline comparison point in genomic compression research and literature.}

We evaluate \emph{end-to-end throughput} based on the throughput of data preparation and genome sequence analysis. We use the read mapping throughput reported by the original paper~\cite{chen2023gem}. 
For data preparation, we use a high-end server with 128 physical cores, as detailed in \sect{\ref{sec:methodology}}. 
We use the best-performing thread count (i.e., after which adding more threads does not improve performance).
I/O operations (reading compressed data), decompression, and read mapping operate in a pipelined manner and in batches, which enables \emph{partial} overlapping \oii{of} these \oii{three} steps. Note that the decompressed data batch \oii{is} stream\oii{ed} directly to the \oii{read} mapper, without being written to the SSD. Further methodology details are in \sect{\ref{sec:methodology}}.

\head{Observations} \fig{\ref{fig:motivation-overhead}} shows the end-to-end throughput normalized to the configuration with \nspring. 
\nh{We observe that data preparation greatly \oii{limits} end-to-end performance. \oii{If the data preparation bottleneck is eliminated, there would be 12.3$\times$ and 4.\oii{0}$\times$ average speedup for \pigz and \nspring.}}

\begin{figure}[h]
  \centering
  \includegraphics[width=0.95\columnwidth]{Figures/motivation-allRS.pdf}
  \caption{\nh{End-to-end throughput for different read sets.}}
  \label{fig:motivation-overhead} 
\end{figure}

\subsection{Challenges and Goal}
\label{sec:motivation-goal}

\nh{\head{Requirements}} Our observations demonstrate that as accelerators dramatically speed up genome sequence analysis, data preparation emerges as a critical bottleneck in the execution pipeline. Note that storing read sets uncompressed is an inefficient and unsustainable way to mitigate this bottleneck, as they are typically 2-40$\times$ larger. Keeping ``hot'' data uncompressed is \emph{not} efficient either because genome analysis often involves many read sets that must be processed with the same priority (e.g., a cancer genomics study of whole-genome read sets can range from tens~\cite{drost2017use} to thousands of samples~\cite{weinstein2013cancer}, each around tens to hundreds of gigabytes compressed). As a result, even the ``hot'' data is too large to be stored efficiently uncompressed. 
Thus, it is common practice to store read sets compressed even when they are \emph{not} in an archival state, and to support integration with decompression (as seen in widely-used genome analysis software, e.g.,~\cite{li2018minimap2,berger2023navigating}).

\oii{Motivated by our observations, we conclude that} there is a critical need to effectively mitigate the data preparation bottleneck while meeting three key requirements: \inum{i}~\emph{high performance and energy efficiency}, \inum{ii}~\emph{high compression ratios}, comparable to state-of-the-art genomic compressors, and \inum{iii}~being \emph{lightweight} to enable seamless integration with a broad range of genome sequence analysis systems  (e.g., in FPGAs, GPUs, ASICs, portable devices, or NDP).

\head{Challenges} \nh{Meeting all three requirements is challenging.} This \oii{challenge is even larger in} genome analysis systems \oii{used} in hardware resource-constrained environments. Examples include portable genomics devices (e.g.,\oii{~\cite{MinIONMk1CO,palatnick2020igenomics,Ballard2018,Oehler2023,watsa2020portable,Quick2016,wang2021nanopore}}), which are critical to facilitating the wide adoption of genomics, or various NDP accelerators (e.g.,\oii{~\cite{wu2021sieve,shahroodi2022krakenonmem,shahroodi2022demeter,dashcam23micro,hanhan2022edam,zou2022biohd, cali2020genasm, huangfu2018radar, khatamifard2021genvom, gupta2019rapid, li2021pim, angizi2019aligns, zokaee2018aligner,Zhang_2023_alignerD,mansouri2022genstore,abakus23taco,megis,jun2016storage,soysal2025mars,kim2025nmp,zheng2025storage,cali2022segram,kim2018grim,kaplan2020bioseal,mao2022genpip,angizi2020pim}}), which lead to large benefits by mitigating data movement overheads of analyzing large-scale genomic data, but are typically implemented in constrained settings (e.g., within the memory system or storage device).

Some works accelerate computational kernels, such as BWT~\cite{qiao2019fpga,guo2013gpu,zhao2017streaming}, FM-index search~\cite{jiang2021exma,arram2015fpga}, or LZMA~\cite{chen2023efficient, leavline2013hardware}, which are widely used in  genomic compressors (e.g.,\ov{~\cite{lan2021genozip,Meng2023,kowalski2019pgrc,chandak2017compression,chen2023efficient}}). 
Despite their benefits, these works \oii{have two limitations. First, they have} high demands for resources like DRAM bandwidth~\cite{zhao2017streaming,wang2018accelerating,dragenora}, on‑chip buffers~\cite{qiao2019fpga}, DRAM capacity~\cite{jiang2021exma}, or compute resources~\cite{arram2015fpga,guo2013gpu,leavline2013hardware}. Efficiently meeting these \oii{hardware resource} demands is challenging, particularly in resource-constrained environments.\footnote{While some genomic (de)compression algorithms\ovi{~\cite{rajarajeswari2011dnabit,saada2016dna}} do not rely on expensive resources, they achieve poor compression ratios, averaging 5.3$\times$ lower than state-of-the-art genomic compressors~\cite{chandak2018spring,Meng2023} on our datasets.} \oii{Second, these works} accelerate only specific kernels and not the \oii{end-to-end} genomic decompression process (e.g.,~reconstructing the full reads from the consensus \oii{sequence} and mismatches). Thus, they do not fully mitigate the data preparation bottleneck (as shown in \sect{\ref{sec:evals}}).

\nh{We use an example to elaborate on the limitations of \oii{resource-intensive data preparation techniques when} integrated with genome analysis systems in resource-constrained environments.}
Consider NDP genome analysis systems implemented inside the resource-constrained environment of the SSD (e.g.,\oii{~\cite{mansouri2022genstore,abakus23taco,megis,jun2016storage,zheng2025storage}}). To benefit from such systems, it is essential to efficiently perform data preparation inside the SSD, since moving the data outside the SSD for preparation \emph{completely undermines} the fundamental benefits of in-storage NDP. Unfortunately, 
performing large amounts of random accesses as needed for data preparation (\oii{matching patterns in large data
structures during decompression}) in the SSD  is inefficient. \oii{This is} due to costly \oii{resource contention within} the SSD (e.g.,~in channels and high-latency NAND flash chips\oii{~\cite{nadig2023venice,tavakkol2018flin,kim2022networked,cho2024aero,kim2025lazy}}). Although modern SSDs have an internal low-latency DRAM buffer, its capacity is small (e.g., 4 GB for a 4-TB SSD~\cite{samsung860pro}),
with over 95\% of it filled with mapping metadata, and its bandwidth is constrained by its \emph{single} channel~\cite{zou2022assasin}. 
\nh{Our motivational analysis (\sect{\ref{sec:motivation-observations}}) emphasizes the \oii{mismatch} between the \oii{available resources} and data preparation's resource demands:} 
across our datasets, we observe that the state-of-the-art genomic decompressors~\cite{chandak2018spring,Meng2023} require random accesses to large amounts of data (up to 26 GB) and with high bandwidth. \oii{Consequently,} even on a high-end system \oii{used in our analysis}, with \emph{eight} DRAM channels, \oii{128 cores, \oiii{and 256 hardware threads,}} the performance of \oii{these genomic decompressors} saturates after \oii{32 threads} due to insufficient main memory bandwidth.

\nh{\textbf{Our goal} is to mitigate the data preparation bottleneck while achieving high performance and energy efficiency, high compression ratios, and a lightweight design. Doing so \oii{would} unlock the full potential of genome sequence analysis acceleration.}

\section{\proposal: Overview}
\label{sec:mech-overview}

We propose \textbf{\proposal}, an algorithm-architecture co-design for highly-compressed \textbf{\underline{s}}torage and high-performance \textbf{\underline{a}}ccess of large-scale \textbf{\underline{ge}}nomic \nh{sequence} data.
\proposal{}
\inum{i}~mitigates the data preparation bottleneck,
\nh{\inum{ii}~achieves high compression ratios, and
\inum{iii}~is lightweight for seamless integration \nh{with a broad range of genome analysis systems}.}
\nh{\proposal is versatile, supporting data from different sequencing technologies and species.} 
\nh{\proposal's approach is based on the \textbf{key insight} that the information encoded in genomic sequence data follows specific trends, shaped by factors such as sequencing technology (e.g., error rates and read lengths) and common genetic phenomena (e.g., typical spatial distributions of genetic variations within genomes). By carefully taking these trends into account when synergistically co-designing algorithms and hardware, \proposal achieves high compression ratios, comparable to state-of-the-art genomic-specific compressors, while enabling \oiii{low} decompression \oiii{latency} using only lightweight hardware and efficient streaming accesses.} \proposal's synergistic co-design is the key enabler of SAGe’s lightweight, high-performance, and energy-efficient design.

\proposal's co-design comprises four aspects. 
First, during compression, \proposal encodes \nh{the} information of reads 
in hardware-friendly, lightweight \emph{data structures} 
and 
exploits genomic data properties in each read set to losslessly \emph{compress}  these structures \textbf{(\sect{\ref{sec:mech-alg}})}.  
Second, we design \emph{lightweight hardware units} \textbf{(\sect{\ref{sec:mech-hw}})} to efficiently interpret and decompress data. 
Third, we design an \emph{efficient data layout} \textbf{(\sect{\ref{sec:mech-ftl}})} to leverage the storage system’s full bandwidth when accessing genomic data. Fourth, we design \emph{specialized interface commands} \textbf{(\sect{\ref{sec:mech-interface}})} that are exposed to genome analysis applications \oiii{to access data and communicate with \proposal{}'s hardware to decompress the data to the
desired format.}

\fig{\ref{fig:sage-hl-overview}(a)} shows \oiii{a} high-level overview of \proposal{}'s data preparation for a genome analysis system. When the genome analysis system requests data using \proposal's interface commands (\circled{1} in \fig{\ref{fig:sage-hl-overview}(a)}), \proposal starts operating based on its \oii{data layout} (\circled{2}).  
\proposal receives compressed data from the storage device (\circled{3}), decompresses it into the desired format using \proposal's hardware  (\circled{4}), and feeds it to the genome analysis system (\circled{5}). \proposal's lightweight design enables it to efficiently integrate with a broad range of genome analysis systems. In \sect{\ref{sec:integration}}, we demonstrate case studies of \oiii{three ways \proposal can} integrate with different genome analysis systems.

\fig{\ref{fig:sage-hl-overview}(b)} shows the high-level overview of how \proposal handles the compression and storage of genome sequence data. 
Since compression is not on the critical path of genome sequence analysis, we perform it on the host system.
The host compresses genome sequence data based on \proposal{}'s compression scheme (\circled{1} in \fig{\ref{fig:sage-hl-overview}(b)}). When writing the \proposal-compressed data, \proposal leverages its customized interface commands (\circled{2}) and data layout to facilitate efficient accesses later, during decompression (\circled{3}).

\begin{figure}[h]
  \centering
  \vspace{0.5em}
  \includegraphics[width=\columnwidth]{Figures/cr-sage-hl-overview.pdf}
  \vspace{0.5em}
  \caption{\oiii{High-level overview of \proposal{}'s (a) data preparation and (b) data compression and storage.}}
  \vspace{0.5em}
  \label{fig:sage-hl-overview} 
\end{figure}

\section{\proposal: Detailed Design}
\label{sec:mech-detailed-design}

\subsection{Compression Algorithm and Data Structures}
\label{sec:mech-alg}

\newcommand\npaa{MMPA\xspace}
\newcommand\npb{MMPGA\xspace}

\begin{figure}[b]
  \centering
  \includegraphics[width=\columnwidth]{Figures/cr-mech-overview-b-q.pdf}
  \caption{Overview of \proposal's compression mechanism.}
  \label{fig:overview} 
\end{figure}

\fig{\ref{fig:overview}} shows an overview of \proposal's lossless compression technique and its hardware-friendly encodings. 
\proposal{} \oiv{exploits} the consensus-based approach (\sect{\ref{sec:background-compression}}) widely used in genomic compression (e.g.,~\cite{chandak2017compression,dufort2021renano,kokot2022colord,kowalski2019pgrc}), storing each read's mismatch information relative to a consensus sequence (\circled{1}).
Similar to existing genomic compressors, \proposal identifies the mismatches during compression by mapping reads to the consensus sequence.\footnote{Note that this is independent from and does not eliminate the need for read mapping in the later genome analysis stages~\cite{Siren2024,Sedlazeck2018}.}
The key difference between \proposal's compression technique over existing genomics-specific compression technique\oiv{s} is the encoding of mismatch information: instead of using expensive backend general-purpose compressors used in genomic-specific compressors (e.g.,\oiv{~\cite{chandak2018spring,Meng2023,lan2021genozip,kowalski2019pgrc,chandak2017compression,chen2023efficient}}), which require costly
computational units, large buffers, or large DRAM bandwidth
or capacity (e.g., due to many random accesses to large data
structures for matching patterns),
\proposal \inum{i}~stores mismatch information in data structures that can be efficiently decoded by lightweight operations and streaming accesses, and \inum{ii}~minimizes their sizes by leveraging read set properties.

To this end, \proposal sequentially stores \oiv{different components of the} mismatch information of \oiv{all} reads in \oiv{the entire} read set contiguously in \oiv{three} arrays. \oiv{These arrays are dedicated to storing different parts of the reads' mismatch information, i.e., \inum{i}~mismatch positions, \inum{ii}~mismatch bases and types, and \inum{iii}~matching positions. \fig{\ref{fig:overview}} shows an example of the array used for storing mismatch positions (\circled{2}).} 
\proposal then adapts the bit width of each array entry and uses a \emph{guide array} to \oiv{indicate the number of bits used for each array entry} (\circled{3}). 
This is effective since the mismatch information in genomic datasets tends to follow specific trends, and by using a limited set of bit widths tailored to these trends, \proposal can significantly reduce data size. 
This approach is lossless since the arrays and guide arrays store a complete and exact representation of all mismatch information, which allows for the perfect reconstruction of each original read's DNA bases from the consensus sequence during decompression.
\nh{In the rest of this section, we discuss how sequencing technology and genetic phenomena influence the trends in each component of mismatch information, i.e., mismatch positions (\sects{\ref{sec:mech-noise-pos-count}), mismatch bases and types (\ref{sec:mech-bases-types}), and matching positions (\ref{sec:mech-map-pos}}). Since these trends vary for each read set, during compression, \proposal tunes the parameters of arrays and guide arrays for \emph{each read set} (\circled{4}). 
The parameters are then encoded at the beginning of the compressed file.} 

The design of the arrays and guide arrays allows them to be interpreted via streaming accesses.
This enables \proposal to reconstruct reads using only lightweight operations during decompression.
As a result, \proposal eliminates the need for expensive hardware resources or frequent random accesses.

\nh{\proposal also performs lossless compression of quality scores (\circled{5}). This feature is optional and can be disabled by the user. This is because some accurate sequencers (e.g.,~\cite{fukasawa2020longqc}) do not report quality scores and print placeholders for format compatibility. 
Given the accuracy improvements in modern sequencing \oiv{technologies}, many workflows (e.g.,~\cite{li2018minimap2,wood2019improved,song2024centrifuger,kolmogorov2019assembly,colquhoun2021pandora}) no longer use quality scores. While some genomic compressors (e.g.,~\cite{Meng2023,vandamme2024tinted,karasikov2020metagraph}) do \emph{not} support quality score (de)compression, \proposal supports optional, lossless quality score (de)compression for broad applicability.}
\nh{\oiv{Since} quality scores do not have the same redundancy patterns
as DNA bases, \proposal compresses quality scores as a separate data stream from DNA bases, a common practice in genomic-specific compressors\oiv{~\cite{chandak2018spring,chandak2017compression,kokot2022colord,roguski2018fastore,kowalski2019pgrc,dufort2020enano,dragenora,chen2023efficient,Deorowicz2020,lan2021genozip,alyami2019lfastqc}}. \proposal maintains the same order for DNA bases and quality scores during compression.}
\nh{\proposal's quality score decompression runs on the host CPU because, as shown in \sect{\ref{sec:mech-qual}}, the throughput of \oiv{\proposal's quality score decompression} is sufficient to avoid becoming a bottleneck in genome analysis pipelines. \sect{\ref{sec:mech-qual}} provides further details about \proposal{}'s quality score compression and decompression.}

\subsubsection{\hm{Mismatch} Positions and Counts}
\label{sec:mech-noise-pos-count}

\begin{figure}[b]
  \centering
  \includegraphics[width=\columnwidth]{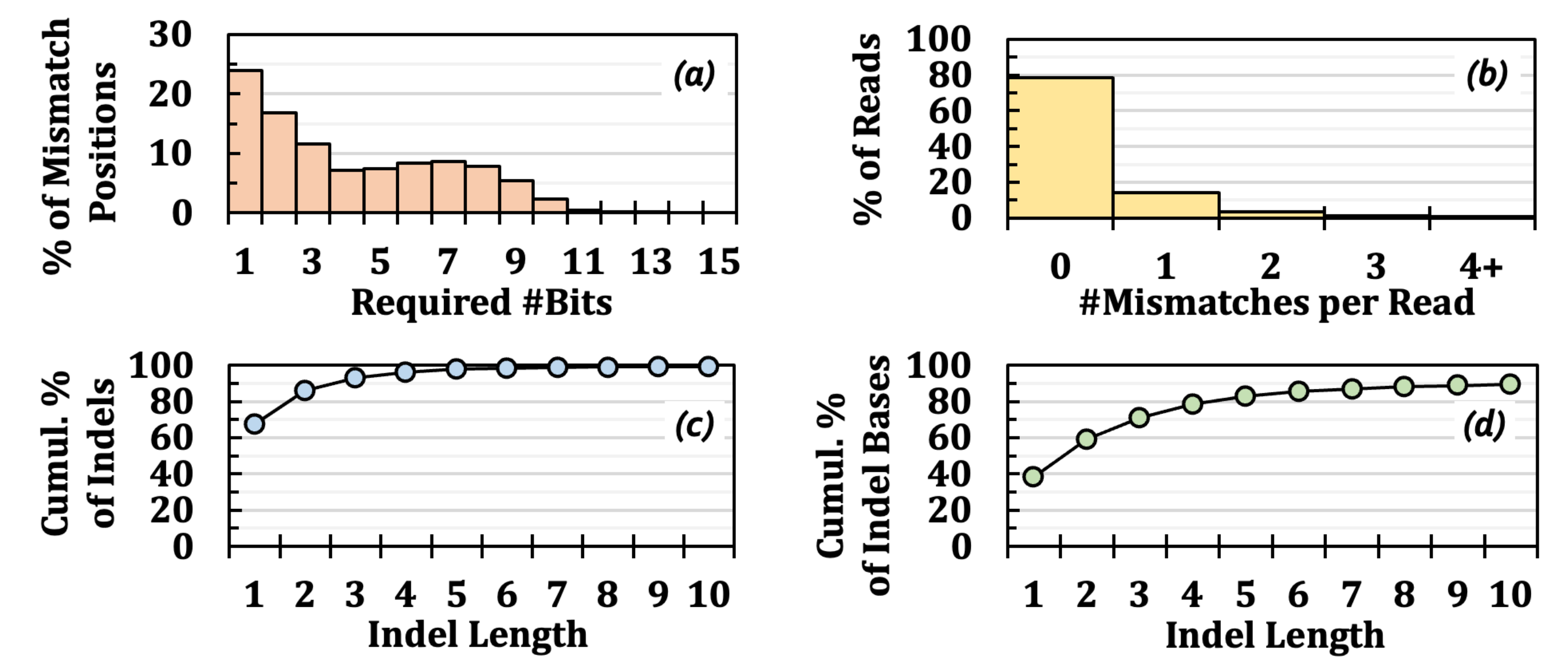}
  \caption{Distribution of  (a)~\#bits needed to store the delta-encoded \hm{mismatch} positions, and (b)~\oiv{mismatch counts per read}. Cumulative distribution
of (c)~indel block lengths, (d)~\#bases \new{in indel blocks}
of different lengths.}
  \label{fig:obsv-noise-pos} 
\end{figure}

\proposal stores \hm{mismatch} positions for reads in a read set sequentially in \oiv{the} Matching Position Array (\emph{\npaa}) and uses \oiv{the} Matching Position Guide Array (\emph{\npb}) to indicate the number of bits used for each \npaa entry and the number of \hm{mismatches} in each read. 
To reduce the sizes of \npaa and \npb, \proposal introduces optimizations based on the properties of read sets.

\fig{\ref{fig:obsv-noise-pos}(a)} shows the distribution of \#bits for the delta-encoded mismatch positions in a representative long read set (RS4 in Table~\ref{table:eval-comp-ratio}). We observe that most delta‑encoded mismatch positions need only a few bits to store\sgprop{1}. This is due to two main factors that can lead to \oiv{spatially} nearby mismatches. First, genetic mutations tend to cluster in some regions of the genome\oiv{~\cite{Bourque2018,Tian2008,amos2013variation}}. Second, a \oiv{regional degradation} in sequencing quality can cause \nh{clusters of} incorrect bases in the read\oiv{~\cite{LaPierre2019,Delahaye2021sequencing,gleeson2021accurate}}. 
Based on this, \proposal performs two optimizations. 
First, during compression of each read set, \proposal tunes the number of distinct bit-counts used for \npaa and the specific values of each bit count. To this end, when finding the mismatch positions of reads, \proposal forms a histogram of bit-counts needed to represent the mismatch positions. \proposal then uses Algorithm~\ref{alg:bitcount} to systematically search for the \oiv{best} bit-count boundaries that minimize the total encoded size for that specific read set based on the information encoded in the histogram.
While the search for bit-counts is exhaustive, its space is limited, i.e.,  constrained by a small upper bound on the required bit-widths (typically $<$ 8 in our evaluation). As evaluated in \sect{\ref{sec:comp-time}}, this makes the optimization cost \oiv{very small}. Second, \proposal exploits the higher frequency of smaller bit counts to further reduce the guide array size by assigning shorter representations to more common inputs\ov{~\cite{huffman2007method}}. For example, assuming four distinct bit counts, \proposal uses variable-length prefix codes 0, 10, 110, and 1110 to represent them (instead of 00, 01, 10, and 11). \oiv{\proposal stores the selected bit counts for the position array and their associated representation in the position guide array in a small \emph{Association Table}.}

\begin{algorithm}[ht]
\color{black}
\caption{Tuning Bit Counts}\label{alg:bitcount}
\begin{algorithmic}[1]
\footnotesize
\Require{Histogram of mismatch position bit counts $H$, where $|H|\leq32$, and a tuning convergence threshold $\varepsilon$.}
\Ensure{List of bit counts $W$ for encoding mismatch positions.}
\Statex
\State $\ell_\text{min}\gets\infty$ \Comment{Initialize min. encoding size}
\State $x_0\gets 0$
\ForAll{$d \in \{1,\ldots,8\}$} \Comment{Find optimal $|W|$.}
  \State $\ell_\text{last}\gets\ell_\text{min}$ \Comment{Best result from previous $d$ values}
  \ForAll{$(x_1,\ldots,x_d)\in|H|^d$ s.t.  $x_0<x_1<\cdots<x_d$}
    \State $\ell\gets$ total length of encoded mismatch positions and guide arrays s.t. positions with bit counts $\in(x_i,x_{i+1}]$ are encoded with $x_{i+1}$ bits.
    \If{$\ell < \ell_\text{min}$}
        \State $\ell_\text{min} = \ell$
        \State $W\gets (x_1,\ldots,x_d)$
    \EndIf
  \EndFor
  \If{$(\ell_\text{last}-\ell_\text{min})/\ell_\text{min} < \varepsilon$}
    \State\textbf{break} \Comment{Exit loop when $\ell_\text{min}$ converges, typically at $d<8$}
  \EndIf
\EndFor\\
\Return{$W$}
\end{algorithmic}
\end{algorithm}

\oiv{Since the number of mismatches can constitute a large fraction of mismatch information in short read sets, we analyze their properties.} \fig{\ref{fig:obsv-noise-pos}(b)} shows \#mismatches per read for a representative short read set (RS2 in Table~\ref{table:eval-comp-ratio}). As also shown by prior works (e.g.,~\cite{nag2019gencache}), most short reads have \hm{no} or few \hm{mismatches} due to short read sequencing's low error rates\sgprop{2}.
We exploit this and use the variable-length encoding \oiv{described} above to assign shorter representations to more common inputs.

Given that \emph{indel blocks} (consecutive inserted or deleted bases) cause many mismatches in long reads, we analyze their properties in a representative read set (RS4 in Table~\ref{table:eval-comp-ratio}).  \fig{\ref{fig:obsv-noise-pos}(c)} and \fig{\ref{fig:obsv-noise-pos}(d)} show the cumulative distributions of indel block lengths and the bases stored for different block lengths, respectively. We observe that \inum{i}~most indel blocks are of length one, a common trait in read sets~\cite{ono2020pbsim2,belyaeva2022best,magi2016characterization,wenger2019accurate,Garg2021}, and \inum{ii}~although single-base blocks are more frequent, larger blocks encompass most indel bases\sgprop{3}. 
\noindent Based on this, we implement two optimizations. First, we store the position of the first \hm{mismatch} and the length of the indel, rather than storing each mismatch position. Second, we use the indel lengths' bit-count distribution to minimize the \#bits needed for their encoding. To this end, we apply the same systematic bit-count tuning  of Algorithm~\ref{alg:bitcount}. Given the strong skew towards 1-bit indels~\cite{ono2020pbsim2,belyaeva2022best,magi2016characterization,wenger2019accurate,Garg2021}, we observe that the \oiv{best} configuration typically consists of two distinct bit counts: one for single-base indels and one for others. Accordingly, after detecting an indel (see \sect{\ref{sec:mech-bases-types}}), we reserve one bit in \npb to indicate whether it is a single-base indel, and otherwise, we dedicate eight bits to encode the indel length, \oiv{as we observe these bit counts to be suitable fits across our evaluated datasets}. In the uncommon case where longer indels are more frequent, \proposal{} can \oiv{use Algorithm~\ref{alg:bitcount} to find other bit count configurations that minimize the encoded size.}

\ov{Based on all the optimizations discussed in this section, \proposal tunes the encoding of position arrays and position guide arrays to compress mismatch position information for all reads in a read set (as described in \fig{\ref{fig:overview}}). Given the lightweight structures of the position arrays and guide arrays, \proposal can decompress mismatch position information with simple operations and efficient access patterns}.

\head{Example Walkthrough \oiv{of Mismatch Position Decompression}} \fig{\ref{fig:example-noise-pos}} shows an example of decompression of mismatch positions, 
demonstrating \oiv{the position array} \npaa (\circled{1}), \oiv{the position guide array} \npb (\circled{2}), and the small \oiv{Association Table}  (\circled{3}). \oiv{Based on the optimizations described in this section, during compression, the bit counts used for representing the \npaa elements are selected, and the associated representations used for these bit counts in the \npb are optimized and stored in the Association Table.} 
In the example of this figure, \oiv{three distinct bit counts are used for mismatch positions stored in the \npaa (i.e., 2, 4, and 8), and these bit count values are represented by 0, 10, and 110 in the \npb.}

To decompress mismatch positions,
\proposal first reads the \hm{mismatch} count of Read\#1 in \npb{} \oiv{(i.e., 0011)}. 
\new{It then scans \npb to obtain the bit counts for each \oiv{three} mismatch position\ov{s} \oiv{of Read\#1} and scans the specified number of bits from \npaa to decode these positions}. 
\oiv{For example, the first entry in \npaa after the mismatch count is 10. Based on the Association Table, \proposal knows that it needs to read the next 4 bits in the \npaa{} (i.e., 1110) to decode the mismatch position.} 
If a \hm{mismatch} is an indel, \proposal checks \npb to determine if the indel is longer than one. If so, it reads the next eight bits from \npaa to determine its length.
After decoding all \hm{mismatch} positions of Read\#1, \proposal proceeds to the next read.

\begin{figure}[h]
  \centering
  \includegraphics[width=0.95\columnwidth]{Figures/cr-noise-pos-example.pdf}
  \caption{Example of decoding \hm{mismatch} positions for long reads in \proposal.}
  \label{fig:example-noise-pos} 
\end{figure}

\subsubsection{\hm{Mismatch} Bases and Types}
\label{sec:mech-bases-types}

\newcommand\nbta{\oiv{M}BTA\xspace}

\begin{figure}[h]
  \centering
  \includegraphics[width=0.8\columnwidth]{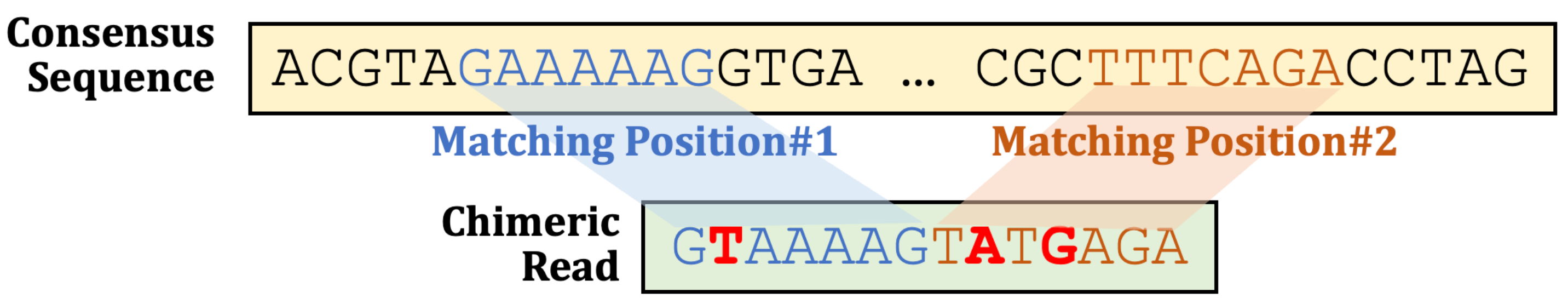}
  \caption{Example of a chimeric read.}
  \label{fig:example-chimeric} 
\end{figure}

\proposal stores \hm{mismatch} bases/types information of reads in a read set in a \hm{Mismatch} Base and Type Array (\emph{\nbta}) and reduces its size with two optimizations based on properties of genomic data. 
First, in long reads, we observe that a significant fraction of \hm{mismatch} bases (e.g., up to 80\% in our datasets) can originate from \emph{chimeric} reads, which are reads with sequences joined from different regions of the genome due to sequencing/library preparation errors, or structural variations~\cite{guan2016structural}\sgprop{4}. 
Parts of these reads map to different locations in the consensus. 
\fig{\ref{fig:example-chimeric}} 
shows a chimeric read with eight mismatches at matching position\#1 and nine at position\#2. 
Thus, considering only the \oiv{top} \hm{matching} position \oiv{(i.e., position\#1 in this example)}, as in prior works (e.g.,~\cite{dufort2021renano,kowalski2019pgrc}), results in many \hm{mismatches} \oiv{(i.e., eight)}. 
Instead of relying on expensive compressors to compress these mismatches, \proposal considers the top $N$ matching positions\footnote{We use N = 3 \oiv{as it led to the best results in our evaluated datasets}. \proposal can tune N to other values as well.} for chimeric reads. 
For example, considering both positions in \fig{\ref{fig:example-chimeric}} \oiv{(positions \#1 and \#2)} reduces mismatches to three, fewer than the \oiv{eight mismatches} at the \oiv{top matching} position.  
Since, in chimeric reads, the number of mismatches at individual positions can be substantial, reconstructing reads from multiple positions (and storing these positions) is more efficient than storing many mismatches from only the top position.

Second, since substitutions are the most common mismatch in short reads due to specifics of sequencing technologies~\cite{glenn2011field,goodwin2016coming,quail2012tale,kchouk2017generations,pfeiffer2018systematic}\sgprop{5}, we design a new encoding to avoid explicitly storing their type.
\proposal compares the mismatching base with the consensus sequence at the corresponding mismatch position. \oiv{If they differ, the mismatch is a substitution.} If they \oiv{are the same}, the mismatch cannot be a substitution and must be an indel. We then use a single bit to distinguish between an insertion and a deletion. This bypasses the need to explicitly store substitution types.

\subsubsection{\hm{Matching} Positions}
\label{sec:mech-map-pos}

\newcommand\mpa{MPA\xspace}
\newcommand\mpb{MPGA\xspace}

Since sequencers read each part of the genome several times (to better allow later analysis to distinguish 
between biological signals and sequencing errors, \sect{\ref{sec:background}}), 
many reads in a read set match closely in the consensus sequence \sgprop{6}. 
Since each read in a read set can be \emph{mapped independently}~\cite{berger2023navigating}, it is possible to reorder them based on their matching position\oiv{s} in the consensus~\cite{chandak2017compression,chandak2018spring,roguski2018fastore,cogo2021genodedup}. Thus, consecutive reads require only a few bits to store their delta-encoded matching positions. 
\oiv{\fig{\ref{fig:obsv-map-pos}} shows this trend by demonstrating the distribution of the number of bits needed to store the delta-encoded matching positions in a representative real human read set (RS2 in Table~\ref{table:eval-comp-ratio}).}
While a similar pattern of \oiv{skew towards specific bit counts} can be seen in other data types (e.g., images~\cite{crisan2021analyzing}), the possibility to \emph{reorder the reads} enables leveraging this pattern in genomic data more efficiently. 
Several genomic compression algorithms (e.g.,~\cite{chandak2017compression,chandak2018spring,roguski2018fastore,cogo2021genodedup}) also reorder reads for delta encoding and then apply entropy coding to further compress the matching positions.
However, their entropy decoding incurs expensive random lookups to match patterns during decompression. Instead, \proposal uses its hardware-friendly structures to significantly compress the matching positions. To this end,
\proposal sequentially encodes the delta-encoded matching positions of reads in a read set in the Matching Position Array (\mpa) and the Matching Position Guide Array (\mpb). \oiv{\proposal then uses the same approach used for reducing the sizes of position arrays and position guide arrays of mismatch positions in \sect{\ref{sec:mech-noise-pos-count}} to reduce the sizes of \mpa and \mpb.}

\begin{figure}[t]
  \begin{center}
    \includegraphics[width=0.85\columnwidth]{Figures/non-wrap.pdf}
  \end{center}
  \caption{Distribution of \#bits needed to store the delta-encoded matching position.}
  \label{fig:obsv-map-pos}
\end{figure}

\ov{\subsubsection{Corner Cases}
\label{sec:mech-corner-cases}

Two corner cases require special handling. First, reads containing unidentified bases (\texttt{N}) expand the DNA alphabet to five characters, 
making 2-bit encoding impossible. Second, some reads include \emph{clips} (i.e., large insertion blocks at the beginning or end) that need reattachment during decompression. 
Adding indicator bits to \ovi{every compressed read to} specify these cases would significantly increase overhead (particularly for short reads). To avoid this, \ovi{\proposal marks a read as a \emph{corner case} by encoding a mismatch as} position~0. \ovi{\proposal} then use\ovi{s} a single bit in the MBTA to indicate whether a read has a mismatch at position~0, indeed, or is a corner case. \ovi{By doing so, \proposal}  efficiently identifies and handles \ovi{corner} cases without impacting standard reads.}

\subsubsection{Quality Scores}
\label{sec:mech-qual}

\nh{\proposal \emph{losslessly} compresses quality scores. As discussed in \sect{\ref{sec:mech-alg}},
this feature is optional and can be disabled.} 
\nh{\proposal compresses quality scores as a separate data stream from DNA bases, a common practice in genomic compressors (\sect{\ref{sec:background}}). \proposal maintains the same order for DNA bases and quality scores when reordering the reads  (\sect{\ref{sec:mech-map-pos}}).}

\proposal's quality score decompression runs on the host CPU because its throughput is sufficient to avoid becoming a bottleneck in genome analysis pipelines. This is because applications that require quality scores, such as variant calling (performed \oiv{on reads generated with sequencing techniques that report quality scores}),  typically only access scores for a small fraction of positions surrounding mismatches (as identified during the earlier read mapping step)\oiv{~\cite{Poplin2018,li2008mapping,Yu2015,Park2025}}. 
For example, across our evaluated read sets, on average only 0.03\% (maximum 10.7\%)\footnote{This ratio \oiv{approaches} upper limits of dissimilarity between an individual's genome and a reference genome (which determines the fraction of accessed quality scores) \oiv{typically tolerated in} genome analysis. At greater dissimilarity levels, \oiv{the efficacy of standard genome analysis is often compromised}~\cite{pearson2013introduction,joudaki2023aligning,prasad2022evaluating}.}
of quality score blocks (with a block size of 25 MB~\cite{grebnov2011libbsc}) are accessed.
This ratio is small because very few genomic locations are variable within a population~\cite{buffalo2021quantifying,10002015global}.
For these quality score blocks, we find that the time to decompress them on the host CPU is significantly shorter than mapping the entire read set using a read mapping accelerator~\cite{chen2023gem}. Therefore, given that reads can be analyzed in batches and in a pipeline\oiv{d manner}, the quality score decompression is \emph{not} on the critical path of the pipeline. \ov{For} our system configuration (\sect{\ref{sec:methodology}}), quality score decompression on the host CPU would not become a bottleneck for cases where up to 17\% of quality scores are accessed. \oiv{This threshold provides a safe margin, ensuring that quality score decompression does not bottleneck execution in the vast majority of cases.}

\proposal's quality score (de)compression is based on the same software (de)compression used for quality scores in \cite{chandak2018spring} (its lossless mode). Since \proposal's quality score decompression runs on the host, it can also flexibly adopt other algorithms \oiv{(e.g.,~\cite{kokot2022colord,Meng2023,dufort2021renano,kowalski2019pgrc,dufort2020enano,dragenora,yang2025gpufastqlz,chen2023efficient,hach2012scalce,roguski2014dsrc2,Deorowicz2020,lan2021genozip,alyami2019lfastqc,cogo2021genodedup})}.

\subsection{Hardware for Data Preparation}
\label{sec:mech-hw}

\newcommand\su{SU\xspace}
\newcommand\rcu{RCU\xspace}
\newcommand\cu{CU\xspace}

\subsubsection{Components}
\label{sec:acc-components}
\fig{\ref{fig:ssdg-acc-overview}} shows the structure of \proposal's lightweight hardware units. 
\proposal consists of three components. \wcirc{1} Scan Unit (\emph{\su}) sequentially scans through \oiv{input position} arrays and \oiv{position} guide arrays to find mismatch information. 
\wcirc{2} Read Construction Unit (\emph{\rcu}) receives the mismatch information from the \su and reconstructs full reads by plugging the mismatches in\hm{to} the correct positions of the \oiv{input} consensus sequence. 
\wcirc{3}~Control Unit (\emph{\cu}) coordinates operations between the \su and the \rcu. 

 \begin{figure}[b]
  \centering
  \includegraphics[width=0.9\columnwidth]{Figures/cr-ssdg-acc.pdf}
  \caption{\oiv{Overview of \proposal's hardware.}}
  \label{fig:ssdg-acc-overview} 
\end{figure}

\proposal enables decompression with streaming accesses, and thus, the \su and the \rcu do \emph{not} rely on large buffers, and instead only require small registers. 
Since \proposal stores mismatches in the order of appearance in each read, decoding is feasible via sequential accesses \oiv{to} the \oiv{position} arrays \oiv{and position guide arrays}. Similarly,  since \proposal stores the reads' positions based on their order in the consensus (similar to other compressors, e.g.~\cite{chandak2017compression,chandak2018spring,roguski2018fastore,cogo2021genodedup}, as explained in detail in \sect{\ref{sec:mech-map-pos}}),
it only sequentially accesses the consensus \oiv{sequence}. 
The \oiv{registers used for buffering the position arrays and position guide arrays} each consist of eight bits since that is the largest element size used in the arrays in our compression scheme \oiv{(\sect{\ref{sec:mech-alg}})}. \oiv{\proposal also uses an eight-bit register} since it loads the configuration parameters in eight-bit chunks. \oiv{\proposal uses a register of size 150~base pairs} since this is the largest read size in most short read sequencing datasets~\cite{glenn2011field,quail2012tale}. For short reads longer than this and for long reads, we reconstruct the reads in 150-base-pair chunks.\footnote{\oiv{As detailed in \sect{\ref{sec:integration}}, in the third integration mode in \fig{\ref{fig:integration}}, \proposal requires two additional 64-bit registers for its operations.}} 
\proposal directly sends reads to the analysis system.
Since reads can be analyzed independently, analysis can start as soon \oiv{\proposal{}'s output} arrives.

\subsubsection{Operations} 
\label{sec:mech-acc-operations}
When \oiv{\proposal receives} a request to access a genomic read set, \oiv{it} starts preparing the reads in the read set. 
The \su reads the tuned configuration parameters of the read set (\circled{1} in \fig{\ref{fig:ssdg-acc-overview}}). 
This includes the \oiv{contents of the Association Table} (as shown in \fig{\ref{fig:example-noise-pos}}).
\oiv{The \su and \rcu then start operating concurrently to decode the mismatch information and reconstruct the full reads.
The \su decodes each read's matching position and mismatch positions by}
scanning through the position guide arrays (\circled{2}) and position arrays (\circled{3}), \oiv{as described in \sect{\ref{sec:mech-alg}}}. Each time a new mismatch position is decoded, the mismatch count \oiv{is} decrement\oiv{ed} (\circled{4}), and when it reaches zero (\circled{5}), the \su reads a new mismatch count (for the next read).
\oiv{Meanwhile, the \rcu decodes the mismatch bases and types (as described in \sect{\ref{sec:mech-bases-types}}) and reconstructs the full reads.}
\oiv{The \rcu scans through the consensus sequence (\circled{6}) and the \nbta (\circled{7})} and when it detects an indel \oiv{mismatch type}, it sends a signal (\circled{8}) to the \su to read the indel length (\circled{9}). 
\oiv{After decoding each read\ov{'}s matching position and its mismatch positions,} the \su sends them to the \rcu (\circled{\small10}). 
\oiv{Based on the information received from the \su, the \rcu reconstructs each read by applying the mismatches to the corresponding positions of the consensus sequence \circled{\small11}, and} 
flexibly formatting the reads (e.g.,~in 2-bit encoded, 3-bit encoded for reads with \texttt{N}, ASCII, etc.) as requested (\circled{\small12}).
\oiv{Finally, \proposal sends the decompressed and formatted reads to the genome analysis system (\circled{\small13})}.

\subsection{Data Layout}
\label{sec:mech-ftl}

\nh{We \oiv{describe} \proposal's efficient data layout, which enables \oiv{it} to leverage the storage system's full bandwidth when accessing and preparing genomic data. We also discuss the feasibility and design efforts required to maintain this layout alongside the existing flash translation layer (FTL).} 
\proposal requires simple changes to the baseline FTL.
\proposal FTL designates each block as genomic or non-genomic.
The SSD recognizes genomic accesses through \proposal commands (\sect{\ref{sec:mech-interface}}). For all other data, vendor-specific FTL features remain untouched, so the SSD behaves like a conventional SSD.

When writing a compressed genomic dataset, \proposal uniformly partitions data across the SSD channels. This is enabled by \proposal's sequential access patterns. 
Each partition of the consensus sequence, along with the compressed mismatch information of the reads matched to that partition, is placed in a separate channel.  
\proposal writes data in a round-robin fashion \oiv{across different channels} such that the active blocks in different channels have the same page offset. This enables \emph{multi-plane} read operations across all channels and leverages the SSD's \emph{full} bandwidth.

During garbage collection (GC)~\cite{tavakkol2018flin,kim2020evanesco,cai2017error,park-dac-2016, park-dac-2019}, we must select victim blocks such that we preserve \proposal{}'s capabilities for multi-plane operations. This requires maintaining the same page offset across all blocks in a parallel unit. Given that genomic read sets do not require partial updates\oiv{~\cite{leinonen2010sequence,cochrane2015international}} and are accessed sequentially, it is straightforward to realize this efficient GC. We perform GC in a grouped manner and select every block in the parallel unit as a group of victim blocks, which are then sequentially rewritten in the order they were originally written, as indicated by their logical address sequence.

\subsection{Interface Commands}
\label{sec:mech-interface}

\proposal in\ov{t}roduces two new interface commands: \oiv{one to request genomic data in the desired format, and another to write compressed genomic data to the storage system}.

\noindent\textbf{\texttt{SAGe\_Read:}} A specialized read command to \inum{i}~access genomic data and \inum{ii}~specify the output format (e.g., 2-bit or 1-hot encoding) required by the genome analysis system. The other parameters (e.g., array/guide array parameters) are written at the beginning of each compressed file and are loaded into the \su (\sect{\ref{sec:mech-hw}}) when accessing genomic data. Upon receiving this command, \proposal operates with its FTL (\sect{\ref{sec:mech-ftl}}). 

\noindent\textbf{\texttt{SAGe\_Write:}} A specialized write command for writing genomic data to SSD and updating its FTL's mapping metadata.

\section{Case Studies of \proposal's Hardware Integration}
\label{sec:integration}

 \begin{figure}[b]
  \centering
  \includegraphics[width=0.9\columnwidth]{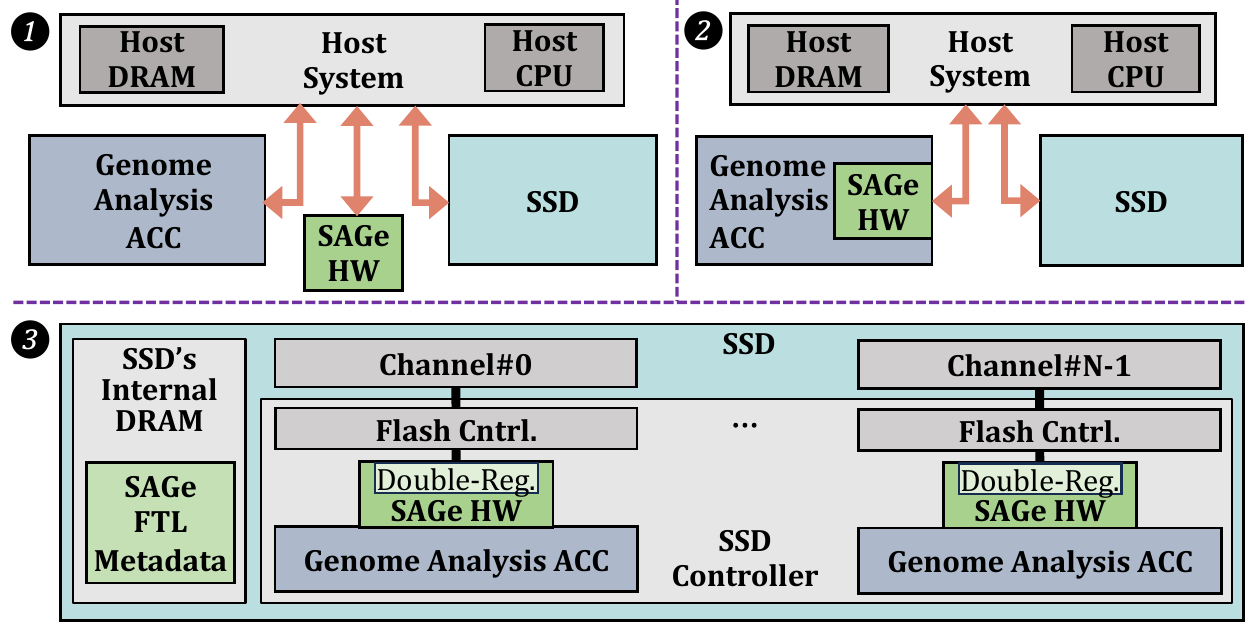}
  \caption{\micro{\proposal's integration with genome analysis systems.}}
  \label{fig:integration} 
\end{figure}

\fig{\ref{fig:integration}} shows three examples of \proposal's integration with genome analysis systems. 
In \circled{1}, 
\proposal{}'s hardware units are connected to the genome analysis system via a PCIe~\cite{PCIE} \oiv{or CXL~\cite{das2024introduction}} interface (i.e., similar to how individual accelerators or GPUs connect). 
In \circled{2}, \proposal's hardware is integrated on the same chip as the genome analysis system (which can be a stand-alone ASIC, or an FPGA, etc.), allowing for tighter integration and saving PCIe \oiv{or CXL \ov{ports},} which are scarce and can be contested by other components (e.g., SSDs, GPUs, or other accelerators). 
This is easily possible due to \proposal's low hardware cost. For example, \proposal's \oiv{logic units (\sect{\ref{sec:eval-area-power}})} consume only 2.5\% of the lookup tables and 0.8\% of the flip-flops of a mid-range FPGA~\cite{fpgamidrange}.
\proposal's lightweight design enables its integration in resource-constrained environments. As an example, in \circled{3}, we demonstrate \proposal's integration with an in-storage NDP genome analysis system (e.g.,~\cite{mansouri2022genstore,abakus23taco,megis,jun2016storage}) on the SSD controller. Given that \proposal performs streaming accesses, \proposal's hardware operates on data fetched from NAND flash chips without needing to buffer them in the SSD's low-bandwidth, single-channel, internal DRAM~\cite{zou2022assasin}.
\proposal performs operations on flash data streams\ov{~\cite{zou2022assasin}} via lightweight double-buffering: one register to hold a chunk of data as computation input, and another to hold the subsequent data chunk arriving from the NAND flash chips. We use two 64-bit registers as they are sufficient to fully utilize the per-channel hardware components and SSD channel bandwidth.

\section{Methodology}
\label{sec:methodology}

\renewcommand\pigz{\texttt{pigz}\xspace}
\renewcommand\nspring{\texttt{(N)Spr}\xspace}
\renewcommand\nsacc{\texttt{(N)SprAC}\xspace}
\newcommand\idec{\texttt{0TimeDec}\xspace}
\newcommand\psw{\texttt{SAGeSW}\xspace}
\newcommand\pmain{\texttt{SAGe}\xspace}
\newcommand\ssdgi{\texttt{SAGe\textsubscript{SSD}}\xspace}

\head{Evaluated Systems}
We evaluate the end-to-end performance of various genome analysis systems, where execution includes \emph{both} data preparation and genome analysis.
For \textbf{data preparation}, we use
\inum{i}~\pigz: \hm{A p}arallel version~\cite{adler2015pigz} of gzip, a commonly-used general compressor. \nh{We exclude other general-purpose compressors from our performance evaluations due to reasons detailed in \sect{\ref{sec:motivation-observations}}.
We do, however, include pigz, i.e., parallel gzip, since it is still widely used as a baseline comparison point in genomic compression research and literature~\oiv{\cite{chandak2018spring,chandak2017compression,kokot2022colord,Meng2023,roguski2018fastore,dufort2020enano,dragenora,yang2025gpufastqlz,chen2023efficient,hach2012scalce,roguski2014dsrc2,Deorowicz2020,lan2021genozip,alyami2019lfastqc,cogo2021genodedup}}};
\inum{ii}~\nspring: Spring~\cite{chandak2018spring} and NanoSpring~\cite{Meng2023}, state-of-the-art compressors for short and long reads, respectively. \pigz and \nspring run on a high-end system, an AMD$^\text{\textregistered}$ EPYC$^\text{\textregistered}$ 7742 CPU~\cite{amdepyc} with 128 physical cores \oiv{(256 hardware threads)}. 
\inum{iii}~\nsacc: \nspring integrated with a BWT-accelerator. There are various accelerators for BWT (e.g.,~\cite{qiao2019fpga,guo2013gpu,zhao2017streaming}). We consider an idealized accelerator that can fully eliminate the BWT execution time from \nspring;
\inum{iv}~\idec: an idealized decompressor (with zero decompression time), but inefficient for integration \micro{in resource-constrained environments (e.g., in our evaluations, for integration with an in-storage NDP genome sequence analysis system)};\footnote{As a  conservative evaluation, \idec can also serve as an idealized representation of the closed-source industry software called ORA~\cite{dragenora}, recently developed for Illumina~\cite{illumina} short read (de)compression. ORA works similarly to other genomic compressors (\sect{\ref{sec:background-compression}}), but uses a different backend compressor for mismatch information, which still suffers from limitations for implementation in resource-constrained environments (detailed in \sect{\ref{sec:motivation-goal}}). 
We cannot use ORA as our short-read decompression baseline due to its closed-source nature, \oiv{but its performance would be upper-bounded by \idec}.}
\inum{v}~\psw: \proposal with its decompression in software, \new{running on the host system (i.e, the same high-end, 128-core system used for \pigz and \nspring), to show the benefits of \proposal's algorithmic optimizations};
\inum{vi}~\pmain: \proposal's full implementation, with its decompression in hardware;
and \inum{vii}~\ssdgi: \pmain with its hardware implemented in the SSD, to integrate with an NDP genome analysis system on the same chip (mode \circled{3} in \fig{\ref{fig:integration}}). 
Decompressors other than \ssdgi connect to the analysis systems using a PCIe interface (\circled{1} in \fig{\ref{fig:integration}}). We do not explicitly \ov{evaluate} mode \circled{2}, \oiv{because given sufficient interface bandwidth, mode \circled{2}}  can perform the same as \circled{1}. As detailed in \sect{\ref{sec:integration}}, \circled{1} enables standalone integration, while \circled{2} integrates on the same chip, thus saving PCIe slots.

We select Spring~\cite{chandak2018spring} and NanoSpring~\cite{Meng2023} as our genomics-specific software baselines because they provide a well-balanced and particularly suitable trade-off in the key aspects of compression ratio, decompression time, losslessness, and being open access. 
In contrast, other tools have less favorable characteristics in compression ratio (e.g.,~\cite{roguski2018fastore,chandak2017compression}), decompression time (e.g.,~\cite{kokot2022colord,dufort2020enano,dufort2021renano}), lossiness (e.g.,~\cite{chandak2017compression,roguski2018fastore}), or do not permit code access and/or modification (e.g.~\cite{dragenora,lan2021genozip}).\footnote{Being able to modify the code is critical for us since we need to adapt it to pass decompressed data batches to the analysis system (without merging data into a single file and writing it to disk).} As explained earlier, we use \idec as a strongly conservative and idealized representation of the closed-source tools.

For \textbf{genome sequence analysis}, we integrate all data preparation configurations with a state-of-the-art read mapping accelerator, GEM~\cite{chen2023gem}. To show \proposal's suitability for resource-constrained environments, we evaluate \proposal{}'s implementation inside the SSD to integrate with a state-of-the-art NDP genome analysis system, GenStore~\cite{mansouri2022genstore}. 
GenStore is an in-storage filter (ISF) that filters reads that do not require expensive read mapping directly inside the SSD, sending only the remaining reads to the mapper, thus alleviating the burden of moving and analyzing a large amount of low-reuse data from the rest of the system (i.e., external I/O, main memory, and compute units). 
\oiv{The resulting pipeline} performs data preparation $\rightarrow$ ISF $\rightarrow$ read mapping. \oiv{The benefits of this pipeline} arise from both \proposal's faster preparation and the ISF operations in \cite{mansouri2022genstore}. The key \oiv{to realizing this pipeline} is that \proposal is the only data preparation configuration that is lightweight enough for efficient implementation inside the SSD. Without \oiv{\proposal}, ISF would require genomic data to be stored uncompressed, which is inefficient, or to decompress data outside SSD, which undermines the fundamental benefits \oiv{of NDP}.

\head{Datasets} We evaluate real-world short- and long-read datasets. \new{Storage systems handle numerous read sets (e.g.,~in a medical center~\cite{Bick2024,Li2023whole} or in a cohort \cite{Bick2024,Li2023whole,10002015global,uk10k2015uk10k}), \hm{so} compressing each read set is essential to reduce the overall burden, regardless of their individual sizes. Thus, we analyze read sets of varying sizes, as \oiv{shown} in Table~\ref{table:eval-comp-ratio}}.

\head{Performance}
We design a simulator that models all components involved during \proposal's execution, including accessing storage, hardware components (both for \proposal and genomic accelerators~\cite{chen2023gem,mansouri2022genstore}), \hm{the} SSD's internal DRAM (for the NDP accelerator~\cite{mansouri2022genstore}), host operations, and their interfaces. \new{Using this methodology, as also demonstrated in prior works (e.g.,~\cite{mansouri2022genstore,park2022flash,megis}), enables us to flexibly incorporate state-of-the-art system configurations in our analysis}. We feed the latency and throughput of each component to this simulator.  
For the components in \textbf{hardware-based} operations,  we implement \proposal's logic \oiv{units} in Verilog and synthesize using Design Compiler~\cite{synopsysdc} at 22~nm~\cite{22gf}.
We use 
Ramulator \oiv{1.0~\cite{kim2016ramulator, ramulatorsource,luo2023ramulator,ramulator2source}} to model \hm{the} SSD's internal DRAM, and MQSim~\cite{tavakkol2018mqsim,mqsimsource} to model the SSD's internal operations.
For the hardware mapper, we use the throughput reported by the original paper~\cite{chen2023gem}.
For the \textbf{software-based} operations (software decompressors), we measure performance with their best-performing thread counts, on a high-end real system, an AMD$^\text{\textregistered}$ EPYC$^\text{\textregistered}$ 7742 CPU~\cite{amdepyc} with 128 physical cores, 256 hardware threads, and 1.5-TB DRAM. 
We perform analysis with both a performance-optimized PCIe SSD~\cite{samsungPM1735} and a cost-optimized SATA SSD~\cite{samsung870evo}. 
Data communication between genome analysis accelerators, \proposal{}'s hardware units, and the storage system is modeled based on the bandwidth of the interfaces between them, depending on integration mode (\fig{\ref{sec:integration}}).
I/O operations, decompression, and genome analysis execute on batches of genomic data in a pipelined manner, which enables partial overlapping \oiv{of} their execution. The synchronization between these stages is modeled via a producer-consumer abstraction.

\head{Prototype Feasibility and Real-System Overheads}
While any real-world prototype will introduce overheads not perfectly captured in simulation, we expect that system-level software changes in the I/O stack to have only a negligible impact. The required changes to the file system and the driver are confined to distinguishing genomics file types and supporting two new interface commands (\sect{\ref{sec:mech-interface}}). Since this adds only simple new logic, we expect it to lead to negligible performance impact. The firmware needs to support \proposal{}’s L2P mapping and GC (\sect{\ref{sec:mech-ftl}}), which are lighter than the baseline (due to \proposal{}'s uniform distribution of data across the SSD channels) and are already modeled by MQSim~\cite{tavakkol2018mqsim}.

\head{Area, Power, and Energy} 
For the accelerator logic units, we use the area and power values obtained from the Design Compiler synthesis~\cite{synopsysdc} of our Verilog HDL implementation of these units in a 22nm technology node~\cite{22gf}. For DRAM, we use the power values of a DDR4 model\oiv{~\cite{ddr4sheet,ghose2019demystifying,ghose2018your}}. For the CPU cores, we use power values obtained from AMD$^\text{\textregistered}$ \textmu{}Prof~\cite{microprof}.  
For \hm{the} SSD, we use the power values of a Samsung 3D NAND SSD~\cite{samsung860pro}. 
We calculate the energy of each component based on its idle and dynamic power and its execution time.
For end-to-end energy, we obtain the energy of the host processor, DRAM, \proposal{}'s logic units, and communication between them.

\section{Evaluation}
\label{sec:evals}

\renewcommand\pigz{\texttt{pigz}}
\renewcommand\nspring{\texttt{(N)Spr}}
\renewcommand\nsacc{\texttt{(N)SprAC}}
\renewcommand\idec{\texttt{0TimeDec}}
\renewcommand\psw{\texttt{SAGeSW}}
\renewcommand\pmain{\texttt{SAGe}}
\renewcommand\ssdgi{\texttt{SAGe\textsubscript{SSD}}\xspace}

\subsection{Performance}
\label{sec:evals-perf}

\begin{figure*}[b]
  \centering
    \includegraphics[width=2\columnwidth]{Figures/cr-eval-new-full.pdf}
        \vspace{-0.5em}
  \caption{End-to-end speedup for different read sets.}
  \label{fig:eval-new-full} 
\end{figure*}

\newcommand\map{\texttt{Map}\xspace}
\newcommand\isfmap{\texttt{ISF}\xspace}

\fig{\ref{fig:eval-new-full}} shows end-to-end \ov{performance}, where execution includes \emph{both} data preparation and genome analysis. 
Speedup is normalized to \nspring.
We make five key observations. 
First, \proposal provides significant speedups. \micro{On the system with the PCIe (SATA) SSD}, \pmain{} leads to 12.3$\times$ \micro{(8.1$\times$)},  3.9$\times$ \micro{(2.7$\times$)}, and 3.0$\times$ \micro{(2.1$\times$)} average speedup over \pigz{}, \nspring, and \nsacc, respectively. 
Second, \pmain{} matches \idec{} in performance because \pmain{} fully hides the decompression overhead in the execution pipeline. As explained in \sect{\ref{sec:motivation}}, I/O, decompression, and read mapping are pipelined, so \ov{overall} throughput depends on the slowest stage. \idec{} and \pmain{} \ov{having} the same performance shows that decompression is no longer the slowest stage.
Third, due to its efficient access patterns, \proposal's implementation in software (\psw) also leads to speedups (2.3$\times$ on average) over \nspring{}. However, \psw's decompression still bottlenecks end-to-end performance. 
\psw{} leads to up to 4.0$\times$ slowdown over \pmain. 
The hardware implementation of \proposal decompression provides greater speedup than its software \ov{version} by handling bitwise operations (on arrays and guide arrays) more efficiently and managing data flow more effectively in a fine-grained manner. 
Ultimately, choosing between these two configurations (software or hardware) is a design decision. 
\proposal software facilitates its ease of adoption in \ov{the} near term, while \proposal hardware 1) provides \emph{better performance}, 2) has \emph{significantly better energy reduction} (across \emph{all} our datasets, as shown in \fig{\ref{fig:eval-energy}}), and 3) is very \emph{lightweight}, which makes it suitable for seamless integration with genome analysis hardware accelerators, even in resource-constrained environments. 
Fourth, by efficiently integrating \ov{\ssdgi{}} with \isfmap{} \ov{(i.e., the in-storage filter implemented inside the resource-constrained environment of the SSD),} \ssdgi{}+\isfmap leads to 7.8$\times$ \micro{(2.5$\times$)} average speedups over \nsacc{} on the system with the PCIe (SATA) SSD. 
\ssdgi{}+\isfmap outperforms \pmain{} in all cases, except when \inum{i}~the input and application do not \hm{largely take advantage} of \ov{\isfmap}{} \new{(i.e., in this case, \isfmap does not filter many reads in the read set)}, and \inum{ii}~\hm{the} SSD's limited external bandwidth bottlenecks performance (e.g., RS1 and RS4 with the SATA SSD). In these cases, \hm{the} \pmain~\new{configuration should be used} to decompress data outside the SSD to avoid moving larger decompressed data through the limited-bandwidth \oiv{storage} interface.
Fifth, regardless of how much decompression tools are optimized for performance, if \ov{their high resource requirements make them} unsuitable for adoption in resource-constrained environments, they miss out on the benefits of a wide range of genome analysis systems \ov{that are implemented in such constrained environments}. For example, \idec{} \ov{(i.e., an idealized decompressor with zero decompression time, but inefficient for integration in resource-constrained environments) cannot \ovi{efficiently and cost-effectively} integrate with \isfmap{} implemented inside the SSD and, as shown in our evaluations, it ends up being} on average 1.8$\times$ (up to  9.9$\times$) slower than \ssdgi{}+\isfmap.

\renewcommand\ssdgi{\texttt{SAGe\textsubscript{SSD}}\xspace}
\newcommand\ssdgo{\texttt{SG\textsubscript{out}}\xspace}

\begin{figure}[h]
  \centering
\includegraphics[width=\columnwidth]{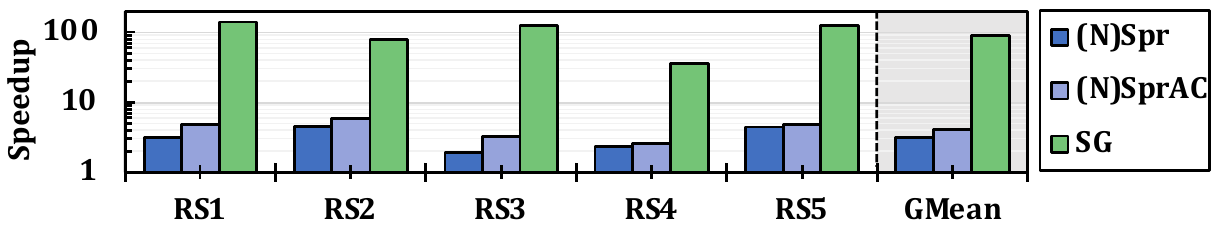}
  \caption{\new{Data preparation speedup.}}
  \label{fig:eval-prep-only} 
\end{figure}

\head{Data Preparation} 
\fig{\ref{fig:eval-prep-only}} shows \emph{only} data preparation throughput, normalized to \pigz{} (on \ov{a} system with the PCIe SSD). \pmain{} leads to 91.3$\times$, 29.5$\times$, and 22.3$\times$ average speedups over \pigz, \nspring, and \nsacc, respectively.

\head{Multiple SSDs}
Since all data in \proposal are accessed in streams, they can be disjointly partitioned between multiple SSDs to enable concurrent access. 
\fig{\ref{fig:multi-ssd}} shows the speedups of \pmain{} and \ssdgi{}+\isfmap over \nspring~in a system with multiple PCIe SSDs. First, \pmain{} maintains its large speedup over \nspring. Second, \ov{with more SSDs,} \ssdgi{}+\isfmap's benefits \ov{increase} for some datasets (RS3 and RS5) since more SSDs improve \isfmap's performance, which was \ov{on} the critical path \ov{of the end-to-end execution}. 
\micro{We conclude that \proposal is able to effectively leverage multiple SSDs. Given this, and since each genomic read can be mapped independently, \proposal's design is also compatible for adoption in distributed storage systems.}

\begin{figure}[h]
  \centering

    \includegraphics[width=0.88\columnwidth]{Figures/cr-eval-multi-ssd-new.pdf}
  \caption{\new{End-to-end speedup with different SSD counts.}}
  \label{fig:multi-ssd} 
\end{figure}

\subsection{Area and Power}
\label{sec:eval-area-power}

Table~\ref{tab:area-power} lists the area and power of \proposal's accelerator units at 1 GHz.
Although they can operate at a higher frequency, their throughput is already sufficient because \proposal's accelerator operations are bottlenecked by the NAND flash read throughput.
\proposal's accelerators consume \hm{a} small area and power of 0.002~mm$^2$ and \ov{0.49}~mW at 22~nm node.

\begin{table}[h]
\centering
\caption{Area and power consumption of \proposal's logic.}
\label{tab:area-power}
\resizebox{\columnwidth}{!}{%
\begin{tabular}{@{\hspace{-0.5pt}}c@{\hspace{-0.02pt}}|c|c|@{\hspace{-0.02pt}}c@{\hspace{-0.05pt}}}
\toprule
\textbf{Logic unit}                  & \textbf{\# of instances} & \textbf{Area [mm\textsuperscript{2}]} & \textbf{Power [mW]} \\ 
\midrule
\midrule
Scan Unit                 &   1 per channel     &   0.000045    &     0.014     \\
Read Construction Unit    &   1 per channel     &   0.000017    &      \ov{0.023}    \\
Double Registers & \multirow{2}*{1 per channel} & \multirow{2}*{0.00020} & \multirow{2}*{\ov{0.035}}\\
(for mode~\circled{3} in \fig{\ref{fig:integration}})    &        &       &          \\
Control Unit              &   1 per channel     &   0.000029    &      0.025    \\
\midrule
\textbf{Total for an 8-channel SSD}           & -                        & \textbf{0.002}    & \textbf{\ov{0.49} (+\ov{0.28} for  mode \circled{3})} \\ \bottomrule
\end{tabular}
\vspace{-0.5em}
}
\end{table}

\subsection{Energy}
\label{sec:evals-energy}

\fig{\ref{fig:eval-energy}} shows the end-to-end energy reduction, \ov{where execution includes \emph{both} data preparation and genome analysis. Energy reduction is}  normalized to \nsacc{} (higher is better). 
We observ\hm{e that} \pmain{} leads to a significant average energy reduction of \nh{34.0$\times$, 16.9$\times$, and 13.0$\times$, over \pigz, \nspring, and \nsacc,} respectively.

\begin{figure}[h]
  \begin{center}
    \includegraphics[width=\columnwidth]{Figures/cr-energy-no-wrap.pdf}
  \end{center}
  \vspace{-0.5em}
  \caption{End-to-end energy reduction.}
  \vspace{-0.7em}
  \label{fig:eval-energy}
\end{figure}

\newcommand\no{\texttt{NO}\xspace}
\newcommand\oone{\texttt{O1}\xspace}
\newcommand\otwo{\texttt{O2}\xspace}
\newcommand\othree{\texttt{O3}\xspace}
\newcommand\ofour{\texttt{O4}\xspace}

\subsection{Compression Ratio}
\label{sec:evals-ratio}

Table~\ref{table:eval-comp-ratio} shows the compression ratios of \pigz, \nspring, and \pmain. For DNA bases, \pmain~achieves 
\inum{i}~2.9$\times$ better average compression ratio than \pigz, 
and \inum{ii}~comparable compression ratios to \nspring, with a modest 4.6\% average reduction.

\begin{table}[h]
\centering
\caption{Compression ratios for different read sets.}
\label{table:eval-comp-ratio}
\resizebox{0.88\columnwidth}{!}{%
\footnotesize
\begin{tabular}{@{\hspace{-0.5pt}}c|@{\hspace{-0.02pt}}c@{\hspace{-0.02pt}}|@{\hspace{-0.02pt}}c@{\hspace{-0.02pt}}|cc|cc|cc}
\hline
\toprule
\multirow{3}{*}{\textbf{Label}} & 
\multirow{3}{*}{\begin{tabular}[c]{@{}c@{}}\textbf{Read Set}\\ (\underline{S}hort, \underline{L}ong)\end{tabular}} & 
\multirow{3}{*}{\begin{tabular}[c]{@{}c@{}}\textbf{Uncomp.}\\ \textbf{Size (MB)}\\ DNA+Qual\end{tabular}} & 
\multicolumn{6}{c}{\textbf{Compression Ratio}} \\
\cline{4-9}
& & & \multicolumn{2}{c|}{\rule{0pt}{2.5ex}PigZ~\cite{adler2015pigz}} & 
\multicolumn{2}{c|}{\rule{0pt}{2.5ex}(Nano)Spring~\cite{Meng2023}} & 
\multicolumn{2}{c}{\rule{0pt}{2.5ex}\proposal} \\
& & & DNA & \multicolumn{1}{@{\hspace{-0.02pt}}c@{\hspace{-0.02pt}}|}{Qual.} & DNA & \multicolumn{1}{@{\hspace{-0.02pt}}c@{\hspace{-0.02pt}}|}{Qual.} & DNA & Qual. \\
\midrule \midrule
RS1 & SRR870667\_2~\cite{Motamayor2013} (S) & 10 000 & 3.39 & 2.23 & 24.8 & 2.80 & 22.8 & 2.80\\
RS2 & ERR194146\_1~\cite{eberle2017reference} (S) & 158 000 & 12.5 & 2.49 & 40.2 & 3.4 & 36.8 & 3.4\\
RS3 & SRR2052419\_1~\cite{zook2016extensive} (S) & 8 000 & 3.41 & 3.45 & 7.2 & 5.07 & 7.1 & 5.07\\
RS4 & PAO89685\_sampled~\cite{ontopendata} (L) & 24 000 & 3.93 & 1.79 & 4.8 & 2.19 & 4.5 & 2.19\\
RS5 & ERR5455028~\cite{Belser2021} (L) & 176 800 & 3.5 & 1.57 & 7.6 & 1.82 & 7.8 & 1.82\\
\bottomrule
\end{tabular}
}
\end{table}

\head{Effect of Input Data Characteristics}
\proposal can support datasets across different species because the consensus sequence is an approximation of the input sample's genomes (\sect{\ref{sec:background-compression}}). 
Given datasets \ov{from} the same species (e.g., RS2 and RS3 in our datasets, which are both \textit{Homo sapiens}), the larger one (R2) achieves higher compression ratios. \ov{Since the storage overhead for the consensus sequence is constant, this one-time overhead is more effectively amortized as the number of compressed reads increases.}

\begin{figure}[h]
  \centering
    \includegraphics[width=0.9\columnwidth]{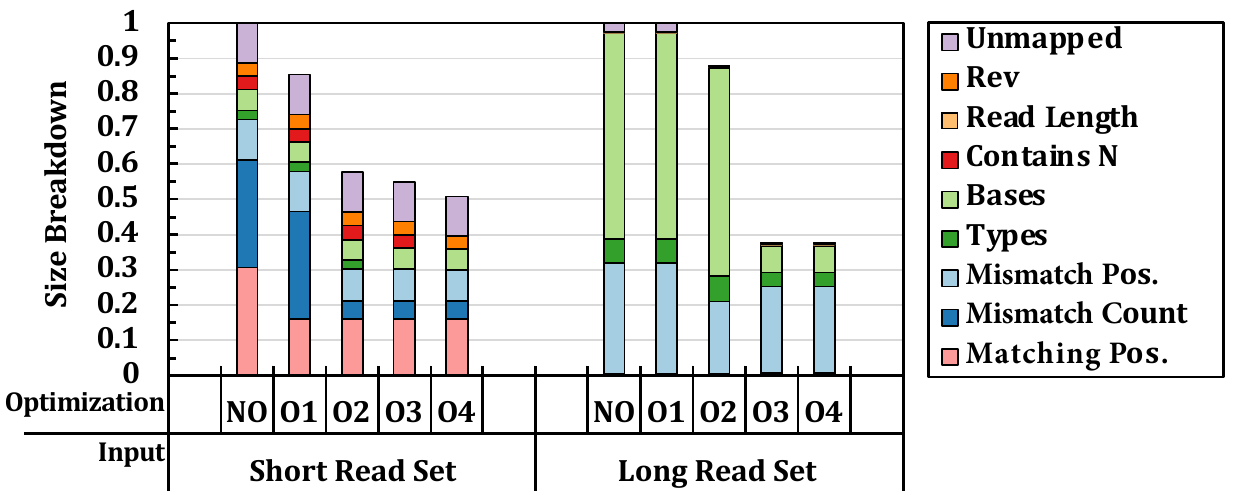}
  \caption{\ov{Effect} of different \proposal optimizations on \ov{the storage size \ovi{of mismatch information}}.}
  \label{fig:eval-step-by-step} 
\end{figure}

\head{\ovi{Effect of Different \proposal Optimizations}} To show the impact of different optimizations, \fig{\ref{fig:eval-step-by-step}}\footnote{\ov{\emph{Unmapped} refers to reads that do not match to the consensus sequence, and} \emph{Rev} refers to a bit marking if a read matches in reverse to the consensus.} presents the size breakdown of the reads' mismatch information for a short (RS2) and long read set (RS4) in five settings:
\inum{i}~\no, with no optimization on the raw mismatch information;
\inum{ii}~\oone, with matching position optimization (\sect{\ref{sec:mech-map-pos}}) added to \no;
\inum{iii}~\otwo, with mismatch positions and count optimizations (\sect{\ref{sec:mech-noise-pos-count}}) added to \oone;
\inum{iv}~\othree, with mismatching base and type optimizations (\sect{\ref{sec:mech-bases-types}}) added to \otwo;
and \inum{v}~\ofour, with corner case optimizations \ov{(\sect{\ref{sec:mech-corner-cases}})} added to \othree. 
We make six observations.
First, \oone significantly reduces the data size of the matching positions in short reads. Note that \oone is not critical for long reads, since matching positions do not constitute a large fraction of mismatch information in longer reads. 
Second, \otwo significantly reduces the data size of mismatch counts in short reads. 
This is because, as detailed in \sect{\ref{sec:mech-noise-pos-count}}\ov{,} most reads in a short read dataset have \hm{no mismatches}, benefiting from \proposal's \hm{mismatch} count encoding.
Third, \otwo leads to a large reduction in the mismatch positions' data size in long reads. 
Fourth, \othree significantly reduces the bases for long reads by efficiently encoding chimeric reads \ov{(\sect{\ref{sec:mech-bases-types}})}. However, the data size for mismatch positions increases in \othree due to the additional mismatches at the chimeric reads' new matching positions. This tradeoff is suitable for \proposal, as it can efficiently encode the greater number of mismatch positions.  
Fifth, \othree reduces the types' data size for short and long reads.
Sixth, \ofour reduces the data size required for labeling corner cases.

\nh{\head{Quality Scores} 
\pmain~achieves 
\inum{i}~33\% better average compression ratio than \pigz,
and \inum{ii}~the same ratios as \nspring. \ov{This is because, as mentioned in \sect{\ref{sec:mech-qual}}, \proposal{}'s quality score (de)compression is based on the same
software (de)compression used for quality scores in \nspring.}}

\subsection{\ov{Resource Requirements}}
\label{sec:eval-context}

Table~\ref{tab:context} \ovi{quantifies} \proposal{}’s lightweightness and effectiveness \ov{by demonstrating the resource requirements of \proposal and other tools}. This table lists the state-of-the-art general-purpose and genomics-specific compression tools implemented on GPUs, FPGAs, ASICs, and CPUs.\footnote{We report compression ratio based on the ratio achieved by each tool’s compression algorithm on \emph{our} datasets. For decompression throughput, we report the best performance reported by the original works.} We make three observations. 
First, leveraging its synergistic, data-aware co-design, \proposal yields the largest average decompression throughput. 
Second, \proposal achieves a significantly higher compression ratio than general-purpose decompressors and a comparable one to genomics-specific decompressors (and higher than the GPU-based genomics\ov{-specific decompressor~\cite{yang2025gpufastqlz}}). 
Third, \proposal eliminates the need for costly and power-hungry computational resources and large memory capacity or bandwidth since, unlike many \ov{de}compression techniques, it avoids frequent random accesses for matching patterns in large data structures.

\begin{table}[t]
\centering
\caption{Comparison \ovi{of} decompression tools.}
\label{tab:context}
\resizebox{0.95\columnwidth}{!}{%
\color{black}
\begin{tabular}{m{8em}||m{4em}|m{3em}|m{2em}|m{11em}|m{5em}|m{6em}}
\toprule
\textbf{Tool}                  & \textbf{Genomics-specific} & \textbf{Avg. Comp. Ratio} & \textbf{End-to-end?} & \textbf{Hardware Requirements} & \textbf{Memory Footprint} & \textbf{Decomp. Throughput (GB/s)}\\
\midrule
\midrule
nvCOMP (DEFLATE)~\cite{nvcomp} & $\times$ & 5.3 & $\checkmark$ & \shortstack[l]{GPU\\ (NVIDIA A100~\cite{nvidiaa100})} & 1.5 GB & 50\\\midrule
Xilinx/AMD GZIP Engine~\cite{amdgzip} & $\times$ & 5.3 & $\checkmark$ & \shortstack[l]{FPGA\\ (AMD Alveo U50~\cite{amdalveou50})} & 80 KB & 0.7\\\midrule
xz~\cite{xz} & $\times$ & 6.7 & $\checkmark$ & \pbox{10em}{CPU\\ (AMD\textregistered{} EPYC\textregistered{}\\ 7742 CPU~\cite{amdepyc}\\ with 128 physical cores)} & 13 GB & 0.6\\\midrule
HW-accelerated ZStandard~\cite{choi2024hardware} & $\times$ & 6.7 & $\checkmark$ & \shortstack[l]{ASIC\\ (1.89 mm$^{2}$ \ov{in a}\\\ov{14nm \ovi{technology} node})} & 2--64 KB & 3.9\\\midrule
GPUFastqLZ~\cite{yang2025gpufastqlz} & $\checkmark$ & 5.8 & $\checkmark$ & \pbox{10em}{GPU\\ (A GPU server with four NVIDIA Tesla V100 GPUs ~\cite{nvidiateslav100})} & N/A\footnote{\ov{We cannot report this value since decompression performance is not reported by the paper and the FPGA implementation is not open-sourced.}} & 7.8\\\midrule
repaq~\cite{chen2023efficient} & $\checkmark$ & 17.1 & $\times$ & \shortstack[l]{FPGA\\ (AMD ALVEO U200~\cite{amdalveou200})} & 16 GB & N/A\\\midrule
Spring~\cite{chandak2018spring} \& NanoSpring~\cite{Meng2023} & $\checkmark$ & 16.9 & $\checkmark$ & \pbox{10em}{CPU\\ (AMD\textregistered{} EPYC\textregistered{}\\ 7742 CPU~\cite{amdepyc}\\ with 128 physical cores)} & 26 GB & 0.7\\\midrule
\proposal & $\checkmark$ & 15.8 & $\checkmark$ & \shortstack[l]{ASIC\\
(0.002 mm$^{2}$ \ov{in a}\\\ov{22nm \ovi{technology} node})} & \ov{128} B & 75.4\\
\bottomrule
\end{tabular}
}
\end{table}

\subsection{Compression Time}
\label{sec:comp-time}

Compression of genomic data is typically an offline and one-time task and is \emph{not} in the critical path of genome analysis. For completeness, \fig{\ref{fig:eval-comp-time}} reports the compression times \ov{of \pigz, \nspring, and \pmain}. In all cases, compression is performed on the host CPU. We make two observations. First, \proposal's compression is slightly faster than \nspring. Compression time\ov{s} of both \pmain{} and \nspring{} \ov{are} mainly dominated by finding mismatch information. After this step, \nspring{} uses back-end \ov{general-purpose} compression techniques to compress the mismatch information, while \pmain{} uses the procedure discussed in \sect{\ref{sec:mech-alg}} to 
compress the mismatch information.
Second, genomics compressors (both \nspring{} and \pmain{}) have much longer compression times than general-purpose compressors due to the need to find mismatches. As mentioned in \sect{\ref{sec:background-compression}}, mismatch information enables genomics compressors to 
achieve substantially better compression ratios by
finding long-range similarity patterns that cannot be captured by general compressors.

\begin{figure}[h]
  \centering
    \includegraphics[width=0.87\columnwidth]{Figures/cr-eval-comp-time.pdf}
    \vspace{-0.5em}
  \caption{Normalized compression time.}
  \label{fig:eval-comp-time} 
  \vspace{-0.5em}
\end{figure}

\section{Related Work}
\label{sec:related}

To our knowledge, \proposal is the \emph{first} system to mitigate the data preparation bottleneck in genome sequence analysis. \ovi{\proposal} achieves compression ratios comparable to state-of-the-art genomic compressors and maintaining a lightweight design for integration with a wide range of genome analysis systems.

\head{Accelerating genome sequence analysis}
Many prior works accelerate genome sequence analysis via algorithmic optimizations (e.g.,~\cite{zhang2000greedy,slater2005automated,li2018minimap2,myers1999fast,marco2021fast,marcosola2023optimal,grootkoerkamp2024apa2}),
or hardware acceleration  (e.g.\oii{~\cite{mutlu2023accelerating,alser2020accelerating,alser2022molecules,lou2020helix,lou2018brawl,shahroodi2023swordfish,saavedra2020mining,markus2020benchmarking,subramaniyan2021accelerated,huangfu2018radar,khatamifard2021genvom, cali2020genasm, gupta2019rapid,li2021pim,angizi2019aligns,zokaee2018aligner,turakhia2018darwin, fujiki2018genax, madhavan2014race,cheng2018bitmapper2,houtgast2018hardware,houtgast2017efficient, zeni2020logan,ahmed2019gasal2,nishimura2017accelerating,de2016cudalign,liu2015gswabe,liu2013cudasw++,liu2009cudasw++,liu2010cudasw++,wilton2015arioc,goyal2017ultra,chen2016spark,chen2014accelerating,chen2021high,fujiki2020seedex, banerjee2018asap,fei2018fpgasw,waidyasooriya2015hardware,chen2015novel,rucci2018swifold,haghi2021fpga,li2021pipebsw,ham2020genesis,ham2021accelerating,wu2019fpga,cali2022segram,kim2018grim,kim2025nmp,doblas2025smx,zhang2000greedy,slater2005automated,%
jia2011metabing,kobus2021metacache,wang2023gpmeta,kobus2017accelerating,su2013gpumetastorms,%
Yano2014,Gamaarachchi2020_f5c,%
zhang2023genomix,cervi2022metagenomic,shih_haru_2023,%
liyanage2023efficient,%
wu2021sieve,shahroodi2022krakenonmem,shahroodi2022demeter,dashcam23micro,hanhan2022edam,zou2022biohd,%
Zhang_2023_alignerD, mansouri2022genstore,abakus23taco,megis,jun2016storage,%
singh2024rubicon,xin2013accelerating,firtina2024aphmm,xin2015shifted,alser2017gatekeeper,alser2019shouji,alser2017magnet,alser2020sneakysnake,bingol2021gatekeeper,xin2016optimal,mao2022genpip,kaplan2020bioseal,angizi2020pim}}).
\proposal is orthogonal to these works and can be flexibly integrated with them to further improve their benefits.

\head{Genomics-Specific Compression}
Many works (e.g.,\omcr{~\cite{chandak2018spring,Deorowicz2020,lan2021genozip,alyami2019lfastqc,kowalski2019pgrc,roguski2018fastore,chandak2017compression,cogo2021genodedup,Meng2023,kokot2022colord,dufort2020enano,dufort2021renano,karasikov2022lossless,vandamme2024tinted,dragenora,yang2025gpufastqlz,chen2023efficient,hach2012scalce,roguski2014dsrc2}}) propose genomic compression.
\new{As detailed in \sect{\ref{sec:motivation-goal}}}, 
some works (e.g.,~\cite{guo2013gpu,qiao2019fpga,zhao2017streaming,jiang2021exma,arram2015fpga,wang2018accelerating,lim2025bancroft,chen2023efficient,leavline2013hardware}) accelerate certain computational kernels 
\js{widely} used in genomic compressors. 
Despite their benefits, \ov{these approaches} 
\inum{i}~are unsuitable for resource-constrained environments and/or \inum{ii}~do \emph{not} fully \omcm{mitigate} the end-to-end preparation \omcm{bottleneck} (evaluated in \sect{\ref{sec:evals}}).
\new{While some works\ovi{~\cite{rajarajeswari2011dnabit,saada2016dna}}
use algorithms that do not rely on expensive resources, they compress poorly (\sect{\ref{sec:motivation-goal}}\hm{)}.}
\micro{As a strongly conservative evaluation, \texttt{0TimeDec} (in \fig{\ref{fig:eval-new-full}}) can serve as an idealized representation of decompressors that, despite their optimized performance, cannot be integrated in resource-constrained environments. As shown in \sect{\ref{sec:evals-perf}}, \ov{by integration in a resource-constrained environment, \ssdgi{}+\isf{}} leads to significant \ov{performance} benefits over \texttt{0TimeDec}.}

\head{General-Purpose Compression} Many works propose general\ovi{-purpose} compression in software (e.g.,~\cite{collet2018zstandard,pavlov20167,Brotli,Katz1991US5051745A,goyal2021dzip,goyal2018deepzip,chen2024ha}) or hardware (e.g.,\ovi{~\cite{bartik2015lz4,liu2018data,fowers2015scalable,chen2021fpga,angerd2022gbdi,gao2024beezip,karandikar2023cdpu,9499902,abali2020data,pekhimenko2016case,pekhimenko2015energy,pekhimenko2013linearly,pekhimenko2012base,buyuktosunoglu2024enterprise,ekman2005robust,arelakis2014sc2,vijaykumar2015case}}). However, as analyzed in \sect{\ref{sec:background-compression}}, \ov{general-purpose compressors} 
achieve poor compression ratios for genomic data~\cite{hernaez2019genomic,zhu2013high}.

\section{Conclusion}
\label{sec:conclusion}

We propose \proposal, an algorithm-system co-design for \ov{highly-compressed} storage and \ov{high-performance} access of \ov{large-scale} genomic data, to \omcm{mitigate the data preparation
bottleneck \ov{in genome analysis}}. 
\ov{We leverage properties of genomic data to co-design \proposal{}'s
algorithm and architecture, such that highly-compressed data can be efficiently interpreted by lightweight hardware
and rapidly prepared for analysis.}
\ov{Due to its lightweight design, \proposal can be seamlessly integrated with a broad range of genome analysis systems to mitigate their data preparation bottlenecks and unlock the full potential of genome analysis acceleration.}
\ov{Our evaluations show that \proposal significantly improves the end-to-end performance and energy efficiency of state-of-the-art
genome sequence analysis accelerators, compared
to when the accelerators rely on state-of-the-art software
and hardware decompression tools.}
\ovi{We conclude that addressing the data preparation bottleneck is crucial for unlocking the full potential of genome analysis acceleration, and hope that \proposal will inspire new studies and designs in this relatively unexplored research area.}

\section*{Acknowledgments}

We thank the anonymous reviewers of HPCA 2025, ISCA 2025, MICRO 2025, and HPCA 2026 for feedback. We thank the SAFARI group members for feedback and the stimulating, \ov{inclusive,} intellectual, \ov{and scientific} environment. We acknowledge the generous gifts and support provided by our industrial partners, including Google, Huawei, Intel, Microsoft, and VMware. This research was partially supported by European Union’s Horizon Programme for research and innovation under Grant Agreement No. 101047160 (project BioPIM), the Swiss National Science Foundation (SNSF), Semiconductor Research Corporation (SRC), the ETH Future Computing Laboratory (EFCL), and the AI Chip Center for Emerging Smart Systems Limited (ACCESS). \ov{Jisung Park was supported by the NRF (RS-2025-00519994) and the IITP (RS-2024-00459026) of Korea.
}

\bibliographystyle{unsrt}
\bibliography{refs}

\end{document}